\documentclass[12pt]{elsart}
\makeatletter
\def\ps@copyright{}
\makeatother

\usepackage[a4paper, height=220mm, width=155mm]{geometry}
\usepackage{graphicx}
\usepackage{epsfig}
\usepackage{amsmath}
\usepackage{amssymb}
\usepackage{amsfonts}
\usepackage{mathrsfs}

\newcommand{\eVq}{\ensuremath{\text{eV}^2}}
\newcommand{\Dmq}{\Delta m^2}
\newcommand{\Sol}{\text{sol}}
\newcommand{\Atm}{\text{atm}}

\renewcommand{\>}{\rangle}

\begin{document}

\begin{frontmatter}

\begin{flushright}
    {IFT-UAM/CSIC-07-62} \\
    {MPP-2007-174}
\end{flushright}

\title{$\theta_{13}$, $\delta$ and the neutrino mass hierarchy at a $\gamma=350$ double baseline Li/B $\beta$-Beam } 

\author[ad:Madrid]{Pilar Coloma,}
\author[ad:Madrid]{Andrea Donini,}
\author[ad:Madrid,ad:Munich]{E. Fern\'andez-Mart\'{\i}nez,}
and
\author[ad:Madrid]{J. L\'opez-Pav\'on}

\address[ad:Madrid]{%
  Instituto F\'{\i}sica Te\'orica UAM/CSIC, Cantoblanco, E-28049 Madrid, Spain}

\address[ad:Munich]{%
Max Planck Institut f\"ur Physik, F\"oringer Ring, 6. M\"unchen, D-80805 Germany}

\begin{abstract}
      We consider a $\beta$-Beam facility where $^8$Li and $^8$B ions are accelerated at 
      $\gamma = 350$, accumulated in a 10 Km storage ring and let decay, so as to produce intense $\bar \nu_e$ and $\nu_e$ beams. 
      These beams illuminate two magnetized iron detectors located at $L \simeq 2000$ Km and $L \simeq 7000$ Km, respectively. 
      The physics potential of this setup is analysed in full detail as a function of the flux.
      We find that, for the highest flux considered ($10 \times 10^{18}$ ion decays per year per baseline),
      the sensitivity to $\theta_{13}$ reaches $\sin^2 2 \theta_{13} \geq 1 \times10^{-4}$; 
      the sign of the atmospheric mass difference can be identified, regardless of the true hierarchy, 
      for $\sin^2 2 \theta_{13} \geq 3 \times10^{-4}$;
      and, CP violation can be discovered in $70 \%$ of the $\delta$-parameter space for 
      $\sin^2 2 \theta_{13} \geq 10^{-3}$, having some sensitivity to CP violation down 
      to $\sin^2 2 \theta_{13} \geq 2 \times 10^{-4}$ for $|\delta| \sim 90^\circ$. 
\end{abstract}

\vspace*{\stretch{2}}
\begin{flushleft}
\end{flushleft}

\end{frontmatter}

\section{Introduction}

The results of solar~\cite{Cleveland:1998nv}-\cite{Ahmed:2003kj},
atmospheric~\cite{Fukuda:1998mi,Ambrosio:2001je},
reactor~\cite{Apollonio:1999ae}-\cite{Eguchi:2002dm} 
and accelerator~\cite{Ahn:2002up}-\cite{Michael:2006rx} 
neutrino experiments show that flavour mixing occurs
not only in the hadronic sector, as it has been known for long, but in
the leptonic sector as well.  
The experimental results point to two very distinct mass-squared
differences, $\Dmq_\Sol \approx 7.9 \times 10^{-5}~\eVq$ and
$|\Dmq_\Atm| \approx 2.4 \times 10^{-3}~\eVq$. 
At present, only two out of the four parameters of the three-family leptonic mixing
matrix $U_\text{PMNS}$~\cite{Pontecorvo:1957cp}-\cite{Gribov:1968kq} 
are known: $\theta_{12} \approx 34^\circ$ and $\theta_{23} \approx 43^\circ$~\cite{GonzalezGarcia:2007ib}.  
The other two parameters, $\theta_{13}$ and $\delta$, are still unknown: for the mixing angle
$\theta_{13}$ direct searches at reactors~\cite{Apollonio:1999ae}-\cite{Boehm:2001ik} and three-family global analyses of
the experimental data give the upper bound $\theta_{13} \leq 11.5^\circ$; for the leptonic CP-violating phase $\delta$ we
have no information whatsoever (see, however, Ref.~\cite{GonzalezGarcia:2007ib}).
We have no clue on the ordering of the neutrino mass eigenstates, either, i.e. on the sign of 
the atmospheric mass difference $\Dmq_\Atm$ (of which only the absolute value has been measured). 
It must be stressed that neutrinos being
hierarchically ordered ($\Dmq_\Atm > 0$) or with an inverted hierarchy ($\Dmq_\Atm < 0$)
makes a big difference for cosmology and neutrino-less double $\beta$-decay experiments \cite{boh}.

The full understanding of the leptonic mixing matrix constitutes, together with the discrimination 
of the Dirac/Majorana character of neutrinos and with the measurement of their absolute mass scale, 
the main neutrino-physics goal for the next decade. 
In the recent past, most of the experimental breakthroughs in neutrino physics have been achieved
by exploiting the so-called ``disappearance channels'': by observing a deficit in the neutrinos that reach 
the detector with respect to those expected to be emitted from the source, a positive and eventually unambiguous 
signal of neutrino oscillations has been established. 
New-generation experiments have been proposed to look for the fleeting 
and intimately related parameters $\theta_{13}$ and $\delta$ through ``appearance channels'' 
such as $\nu_e \leftrightarrow \nu_\mu$ (the ``golden channel'') \cite{Cervera:2000kp}
and $\nu_e \to \nu_\tau$ (the ``silver channel'') \cite{Donini:2002rm}. However, strong correlations between $\theta_{13}$ 
and  $\delta$ \cite{BurguetCastell:2001ez} and the presence of parametric degeneracies in the 
($\theta_{13},\delta$) parameter space, \cite{Minakata:2001qm}-\cite{Barger:2001yr}, make 
the simultaneous measurement of the two variables extremely difficult. 
A further problem arises from our present imprecise knowledge of atmospheric parameters, whose uncertainties are far too large 
to be neglected when looking for such tiny signals as those expected in appearance experiments \cite{Donini:2005rn}.

Most of the proposed solutions to these problems imply the combination of different experiments and facilities, such as reactors
(Double Chooz \cite{Ardellier:2006mn} should start data taking in 2008), Super-Beams 
(of which T2K \cite{Itow:2001ee} is the first approved one), $\beta$-Beams \cite{Zucchelli:sa}  
or the Neutrino Factory \cite{Geer:1997iz,Apollonio:2002en}.
To compare different options and to understand which of them is the one to be pursued as a future neutrino facility, a list of observables to 
be measured has been defined: $\theta_{13}$; the CP-violating phase $\delta$;  the sign of the atmospheric mass difference (hereafter called $s_{atm}$); 
the deviation from $\theta_{23} = 45^\circ$; the $\theta_{23}$-octant  (if $\theta_{23} \neq 45^\circ$). 
Aside from these measurements, a new facility should also reduce the present errors on atmospheric and solar parameters. 

A comparison of all the proposed facilities based on this ``shopping list'' has been presented in the  
{\em International Scoping Study of a future Neutrino Factory and SuperBeam facility} (ISS) Report 
\cite{Group:2007kx}. The outcome of this comparison is that the ``ultimate'' neutrino oscillation experiment 
is a Neutrino Factory with 20-30 GeV stored muons, whose (anti)neutrino fluxes aim at two 50 Kton magnetized iron detectors
located at $L \in [2000,4000]$ Km and $L \sim 7000$ Km from the source, respectively. The goal luminosity for such facility 
is $1 \times 10^{21}$ useful muon decays per year per polarity per baseline. The nearest competitor of this setup is a high $\gamma$
($\gamma \in [300,600]$) $^6$He/$^{18}$Ne $\beta$-Beam aiming at a 1 Mton water \v Cerenkov detector located at $L = 650$ Km from 
the source. The nominal luminosity of this second facility is $2.9 \times 10^{18}$ $^6$He ($1.1 \times 10^{18}$ $^{18}$Ne) useful ion decays
per year. 

In this paper we consider the same physics case of these two setups, namely, $\theta_{13} \leq 3^\circ$ and out of 
the reach of the next generation of approved experiments. We propose a setup based on an intense $^8$Li/$^8$B $\beta$-Beam 
accelerated at high $\gamma$, $\gamma = 350$, aiming at two magnetized iron detectors
located at $L = 2000$ Km and $L = 7000$ Km, respectively. We have studied the performance of this setup
as a function of the flux. The outcome of our analysis is that the physics reach of such a facility, 
according to the ``shopping list'' defined above, is comparable with that of the two reference Neutrino Factory--based
and $\beta$-Beam--based setups defined above if a flux of $10 \times 10^{18}$ $^8$Li/$^8$B ion decays per year per baseline
could be achieved.

The paper is organized as follows: in Sect.~\ref{sec:pheno} we present the physics case for the setup proposed in this paper; 
in Sect.~\ref{sec:det} we describe the neutrino fluxes, the $\nu N$ CC cross-sections, the detector, the signal and the expected backgrounds; 
in Sect.~\ref{sec:results} we present the physics performance; in Sect.~\ref{sec:concl} we eventually draw our conclusions.
In App.~\ref{sec:app} we review the accelerator-related aspects of our setup discussing, when possible, the main differences with 
respect to ``standard'' $\beta$-Beams.

\section{The physics case for a $\gamma=350$ double baseline Li/B $\beta$-Beam}
\label{sec:pheno}

\begin{figure}[t!]
\vspace{-0.5cm}
\begin{center}
\begin{tabular}{cc}
\hspace{-0.55cm} \epsfxsize12cm\epsffile{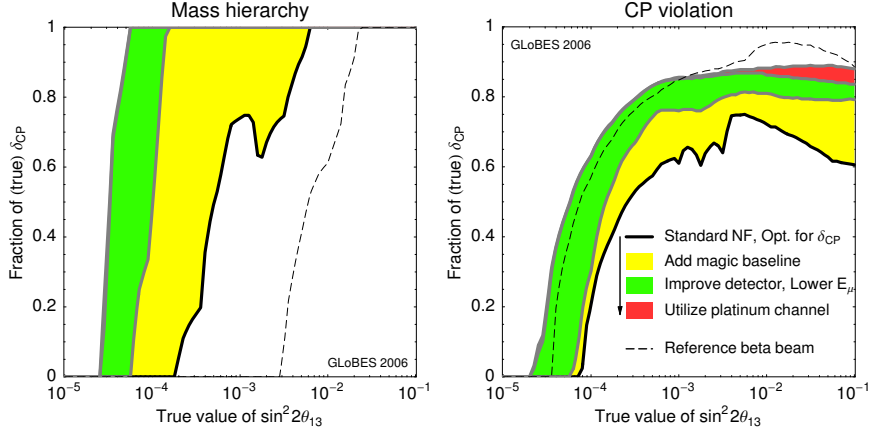}
\end{tabular}
\caption{\it 
Comparing performances at the ISS: the Neutrino Factory baseline setup. >From Ref.~ \cite{Group:2007kx}. 
}
\end{center}
\label{fig:optsummary}
\end{figure}

An example of the physics performances of different setups is shown in Fig.~\ref{fig:optsummary} 
(taken from Ref.~\cite{Group:2007kx}). In the Figure, setups are compared focusing on two of the observables listed above, 
namely the sensitivity to $s_{atm}$, Fig.~\ref{fig:optsummary}(left), 
and to the CP-violating phase $\delta$, Fig.~\ref{fig:optsummary}(right). 
The thick solid line refers to a Neutrino Factory with $1 \times 10^{21}$ useful muons per year per polarity, 
where stored muons have an energy in the range $E_\mu \in [20,30]$ GeV, aiming at a 50 Kton magnetized iron detector 
(whose characteristics have been described in Ref.~\cite{Cervera:2000vy}) located at $L = 4000$ Km from the source. 
This facility is compared with a $\beta$-Beam with $^6$He ($^{18}$Ne) ions accelerated at $\gamma = 350$ $(580)$, aiming at a 1 Mton water 
\v Cerenkov detector located at $L = 650$ Km from the source (dashed line). The (anti)neutrino flux of the latter is
$2.9 \times 10^{18}$ ($1.1 \times 10^{18}$) useful ion decays per year.
This setup was proposed in Ref.~\cite{Burguet-Castell:2003vv,Burguet-Castell:2005va} as a possible alternative to the Neutrino Factory
if $\theta_{13} \leq 3^\circ$, i.e. out of reach for the approved Double Chooz and T2K experiments\footnote{We recall that 
the sensitivity of these experiments to $\theta_{13}$ is $\theta_{13} \geq 5^\circ (\sin^2 2 \theta_{13} \geq 0.03)$ 
for Double Chooz \cite{GilBotella:2007uv} and $\theta_{13} \geq 3^\circ (\sin^2 2 \theta_{13} \geq 0.01)$ for T2K-I \cite{Itow:2001ee}.
In case of a positive signal, new experiments will be launched to look for a CP-violating signal (something out of reach for T2K-I, 
for which only a $\nu_\mu$ flux will be produced) and to improve the precision on $\theta_{13}$. 
Notice that in the experiments of the next generation, no sensitivity to $s_{atm}$ is expected, 
due to the (relatively) short baselines of these experiments (see, however, Ref.~\cite{Mena:2006uw}).}.
In this case only neutrino beams of a new design, such as $\beta$-Beams or the Neutrino Factory,
can observe a signal through the ``golden channel'' $\nu_e \to \nu_\mu$ \cite{Cervera:2000kp}. 

As it can be seen from the Figure, the Neutrino Factory single baseline setup outperforms the high-$\gamma$ He/Ne $\beta$-Beam
when looking for $s_{atm}$, due to its longer baseline. 
However, it is severely outperformed by the latter when looking for $\delta$. This happens because the average neutrino energy 
for NF--based setups is, generally, much above the $\nu_e \to \nu_\mu$ oscillation peak for the considered baselines. 
As a consequence, the ($\theta_{13},\delta)$ intrinsic degeneracy is particularly harmful, resulting in a reduced sensitivity to $\delta$. 
The sign degeneracy is also the cause of a severe loss of sensitivity to $\delta$  for intermediate $\theta_{13}$, $\theta_{13} \in [1^\circ,3^\circ]$, 
see Refs.~\cite{Huber:2002mx,Huber:2006wb}.

This problem is cured in three steps: 
(1) adding a second (identical) detector located 
at the so-called ``magic baseline'' \cite{BurguetCastell:2001ez,Huber:2003ak}, $L \sim 7000$ Km; (2) using new oscillation channels 
(the ``silver'' and ``platinum'' channels \cite{Donini:2002rm,Bueno:2000fg}); and, eventually, (3) with an improved magnetized iron detector,  
with a very good efficiency to select the charge of CC $\nu_\mu$ events with $E_\nu \geq 10$ GeV and a non-vanishing\footnote{The baseline 
detector \cite{Cervera:2000vy} has a vanishing efficiency for CC $\nu_\mu$ events for $E_\nu < 10$ GeV.} efficiency 
for CC $\nu_\mu$ events down to $E_\nu \geq 1$ GeV \cite{cerveragolden07}. 

The three improvements have a similar effect: they help in solving the degeneracy problem in the region of small but not-vanishing 
$\theta_{13}$ (where the problem of parametric degeneracy is more severe). 
After these changes with respect to the single baseline option, the Neutrino Factory setup and the high-$\gamma$ He/Ne $\beta$-Beam
have comparable sensitivity to $\delta$, with the $\beta$-Beam performing slightly better than the Neutrino Factory 
for large $\theta_{13}$, and the Neutrino Factory besting the $\beta$-Beam for small $\theta_{13}$. 
Eventually, the double baseline Neutrino Factory setup improves further its impressive sensitivity to $s_{atm}$ 
with respect to the single baseline option, thanks to the second baseline of $L \sim 7000$ Km. 

After the comparison performed in the framework of the ISS report, the Neutrino Factory Reference Setup (NF-RS) is defined as follows:
a Neutrino Factory with $1 \times 10^{21}$ useful muons per year per polarity per baseline, with $E_\mu \in [20,30]$ GeV; 
the beam is aimed at two 50 Ktons ``improved'' magnetized iron detectors (MIND proposal, as described in Ref.~\cite{cerveragolden07}) 
located at $L \in [1500,4000]$ Km and at the magic baseline, $L \sim 7100$ Km, respectively. 
A variant of this detector includes magnetized emulsions (or liquid Argon) to look for CC $\nu_\tau$ events and take advantage of 
$\nu_e \to \nu_\tau$ or $\nu_\mu \to \nu_\tau$ (see MIND-HYBRID proposal, again in Ref.~\cite{cerveragolden07}). 

This setup is rather expensive and  extremely demanding from the technical point of view
(both things are, of course, highly correlated). In particular, to produce and maintain such an intense neutrino flux, 
state-of-the-art technology is not enough.  
A short-list of technical problems still to be solved to pursue this setup consists (at least) of:
(1) the design of a target that can survive a sustained 4MW proton current 
    (the MERIT R\&D experiment is currently studying this problem, \cite{MERIT});
(2) the amount of muon cooling that is needed to accelerate and fill the storage ring such as to 
       maintain the desired neutrino flux must be evaluated. Different options to cool the muons
       including Linac's, FFAG's or hybrid techniques are under investigation 
       (the MICE R\&D experiment is currently studying the possibility to cool muons through LINAC's, \cite{MICE}, 
       whereas a nested FFAG's chain is under construction in Japan);
(3) the expected radioactivity in the proximity of the target and of the storage ring must be studied
      and, if huge, reduced to acceptable levels. 

Some of these problems are common to alternative options and some are peculiar to NF--based
setups. For example, R\&D on new targets is useful for very intense Super-Beams, also, whereas 
activation of the storage ring is a concern of $\beta$-Beam options, too. A new proposal should be 
compared with the reference setup (as defined above) taking into account at the same time the 
physics performance according to the ``shopping list'', its technical feasibility and its cost. 

We face the same physics case considered for the NF-RS and the high-$\gamma$ He/Ne $\beta$-Beam 
(i.e., $\theta_{13} \leq 3^\circ$) by taking advantage of some of the specific characteristics of the two facilities. 
We consider, thus, a setup based on an intense $^8$Li/$^8$B $\beta$-Beam 
accelerated at high $\gamma$, $\gamma = 350$, aiming at two magnetized iron detectors
located at $L = 2000$ Km and $L = 7000$ Km, respectively.

This proposal is the natural conclusion of a series of theoretical, experimental and accelerator achievements: 
\begin{itemize}
\item
In Ref.~\cite{Zucchelli:sa} the idea of accelerating radioactive ions and store them so as to produce intense $\nu_e (\bar \nu_e)$ beams
was advanced. In the original proposal, $^6$He/$^{18}$Ne ions were boosted at $\gamma \sim 100$ using existing infrastructures at CERN,
producing $\nu_e (\bar \nu_e)$ beams aimed at a 1 Mton water \v Cerenkov detector to be located in a newly excavated cavern 
at the Fr\'ejus underground laboratory, $L = 130$ Km down the source. The physics reach of this setup was studied in 
Ref.~\cite{Bouchez:2003fy,Donini:2004hu,Donini:2004iv}.
\item
In Refs.~\cite{Burguet-Castell:2003vv,Burguet-Castell:2005va} it was proposed to accelerate the same two ions, $^6$He and $^{18}$Ne, 
at a much higher $\gamma$ ($\gamma = 350$ and 580, respectively), aiming again at a 1 Mton detector water \v Cerenkov detector 
to be located at a newly excavated cavern at the Canfranc underground laboratory, $L = 650$ Km from the source. 
Such a high Lorentz boost factor could only be attained at CERN using new infrastructures. A new SPS, the SPS+, is actually under 
discussion in the framework of the planned LHC maintenance and upgrade programme \cite{PAF}. Alternatively, the TeVatron could be used
for the last acceleration stage (see, e.g., Ref.~\cite{Jansson:2007nm}). This setup greatly outperforms the ``low''-$\gamma$ one discussed above
and could compete with NF--based setups in the sensitivity to CP violation.
\item 
In Ref.~\cite{Rubbia:2006pi,Rubbia:2006zv}, the ``ionization cooling'' technique to produce intense $^8$Li and $^8$B beams was proposed 
(the latter being out of reach with standard ISOL-type targets). The feasibility of this method will be studied in full detail 
in the framework of the EURO-$\nu$ Design Study \cite{Euronu}. 
\item 
In Ref.~\cite{Donini:2006dx}, some of the authors of this paper proposed the use of a ``cocktail'' of $^8$Li/$^8$B and $^6$He/$^{18}$Ne 
$\beta$-beams at $\gamma = 100$ (the maximum that can be achieved with existing CERN infrastructures) 
illuminating a 1 Mton water \v Cerenkov detector located at $L = 650$ Km, so as to solve some of the parametric degeneracy 
from which the measurement of ($\theta_{13},\delta$) is afflicted. 
This setup is only useful in the case of large $\theta_{13}$, due to its statistical limitations.
\item
In Refs.~\cite{Donini:2006tt,Donini:2007qt}, the possibility of using a high-$\gamma$ 
$^6$He/$^{18}$Ne $\beta$-Beam illuminating a (MINOS-like) 50 Kton magnetized iron 
detector located at $L = 732$ Km down the source was explored. The existing cavern at the Gran Sasso underground laboratory
could be used for such a compact detector. This setup is also statistically limited, though.
\item
Eventually, in Refs.~\cite{Agarwalla:2006vf,Agarwalla:2007ai} a $\gamma = 350$ $^8$Li/$^8$B $\beta$-Beam illuminating 
a 50 Kton magnetized iron detector (INO \cite{ino}) located at $L = 7100$ Km down the source was proposed,
to take advantage of the resonant matter effects so as to measure $s_{atm}$ for $\sin^2 2 \theta_{13} \geq 10^{-3}$.
\end{itemize}

The main difference of using $^8$Li/$^8$B  instead of $^6$He/$^{18}$Ne ions is that
the end-point energy of the $^8$Li/ $^8$B $\beta$-decays is  $Q_\beta \sim 13$ MeV
(to be compared with $Q_\beta \sim 3.5$ MeV for  $^6$He/$^{18}$Ne). 
With a Lorentz boost factor of $\gamma = 350$, a (relatively) high mean neutrino energy 
in the laboratory frame, $E_\nu \sim 6$ GeV is achievable.
Tuning the neutrino flux to  the $\nu_e \to \nu_\mu$ oscillation peak, therefore, 
the corresponding baseline is considerably longer than that of the  the high-$\gamma$ 
He/Ne $\beta$-Beam. For this reason, a good sensitivity to the mass hierarchy is foreseen. 
A further consequence of having an energetic neutrino flux is that we can safely use
dense detectors with a good muon identification efficiency, as an alternative to the water \v Cerenkov technology.

We will therefore consider the neutrino beam to be aimed at two 50 Kton magnetized iron detectors 
of the MIND-type, located at $L = 2000$ Km (at the oscillation peak) and at the ``magic baseline'', 
$L \simeq 7000$ Km, as in the NF-RS. 

Notice that, since one of the baselines is tuned to the oscillation peak, this setup is expected to suffer  
from a much softer degeneracy problem with respect to NF--based setups (as it was the case for the high-$\gamma$ He/Ne $\beta$-Beam). 
The longer baseline ($L \simeq 7000$ Km) is far from the first oscillation peak but close to the baseline at which the (anti)neutrino oscillation 
probabilities present a resonant enhancement for normal (inverted) hierarchy, thus providing excellent sensitivities to the mass hierarchy 
\cite{Agarwalla:2006vf,Agarwalla:2007ai,Agarwalla:2006gz}. This baseline is also very close to the ``magic baseline''  
\cite{BurguetCastell:2001ez,Huber:2003ak}, at which matter effects cancel the dependence of the oscillation probability on $\delta$. 
The sensitivity to $\delta$ is then lost at this baseline, but combined with the ``on-peak'' oscillation 
data at the nearer detector ($L \simeq 2000$ Km), the ($\theta_{13}$,$\delta$) intrinsic degeneracy
and the sign degeneracy that afflict the shorter baseline can be easily solved, achieving very good 
sensitivities to CP violation, also. 

\section{Signal and backgrounds}
\label{sec:det}

The $\beta$-beam concept was first introduced in Ref.~\cite{Zucchelli:sa}.  
It involves producing a huge number of $\beta$-unstable ions, accelerating them to some reference
energy, and allowing them to decay in the straight section of a storage ring, resulting in a very intense and pure $\nu_e$ or $\bar \nu_e$ beam. 
``Golden'' sub-leading transitions, $\nu_e \to \nu_\mu$ and $\bar \nu_e \to \bar \nu_\mu$, can then be measured through muon observation 
in a distant detector.
The $\beta$-beam concept shares with the Neutrino Factory two main advantages with respect to conventional beams (where neutrinos are obtained via 
pion decay): a) the neutrino flux is pure (for a $\beta$-beam, only $\nu_e$ or $\bar \nu_e$  neutrinos are present in the flux), thus decreasing 
the beam-induced background, and b) the neutrino spectrum can be exactly computed, thus strongly reducing flux systematics.

The neutrino flux per solid angle in a detector located at distance $L$ from the source, aligned with the boost direction of the parent ion 
is~\cite{Burguet-Castell:2003vv}:

\begin{equation}
\left.{d\Phi \over dS dy}\right|_{\theta\simeq 0} \simeq
{N_\beta \over \pi L^2} {\gamma^2 \over g(y_e)} y^2 (1-y)
\sqrt{(1-y)^2 - y_e^2}, 
\label{eq:flux}
\end{equation}
where $0 \leq y={E \over 2 \gamma E_0} \leq 1-y_e$, $y_e=m_e/E_0$
and
\begin{equation}
g(y_e)\equiv {1\over 60} \left\{ \sqrt{1-y_e^2} (2-9 y_e^2 - 8
y_e^4) + 15 y_e^4 \log \left[{y_e \over
1-\sqrt{1-y_e^2}}\right]\right\} \, ,
\end{equation}
where $\gamma$ is the Lorentz boost factor.
In this formula $E_0$ represents the electron end-point energy in the center-of-mass frame of the $\beta$-decay, $m_e$ the electron mass, 
$E$ the energy of the final state neutrino in the laboratory frame and $N_\beta$ the total number of ion decays per year.

The key parameter in the optimization of the $\beta$-beam flux is the relativistic $\gamma$ factor: if the 
baseline is tuned to be at an oscillation peak for $\nu_e \to \nu_\mu$ transitions, the statistics
that can be collected in the detector scales linearly with $\gamma$ \cite{Zucchelli:sa}.
This can be derived from eq.~(\ref{eq:flux}) as follows: in the hypothesis of linear dependence of the total 
neutrino-nucleon CC cross-section on the neutrino energy and for $L/E$ tuned to the $n$-th $\nu_e \to \nu_\mu$ oscillation peak, 
the number of events expected in the far detector located at distance $L$ is: 

\begin{equation}
N_{events} \propto N_\beta \left ( \frac{\Delta m^2}{2 n -1} \right )^2 \frac{\gamma}{E_{cms}}
\label{eq:stats}
\end{equation}

where $E_{cms}$ is the mean neutrino energy in the center-of-mass frame of the $\beta$-decay 
(with $\<E\> = 2 \gamma E_{cms}$).  Applying this formula, the signal statistics in the far detector increases linearly with the boost factor $\gamma$  
and the number of decays per year $N_\beta$, and it decreases linearly with the mean neutrino energy.

\subsection{Choice of $\beta^\pm$-emitters}

Three parameters that are crucial to determine the choice of the optimal $\beta$-emitters are $E_{cms}$, the ion half-life $T_{1/2}$ and $Z$. 
First of all, from eq.~(\ref{eq:stats}), it can be seen that the lower the mean neutrino energy $E_{cms}$, the larger 
the statistics\footnote{Notice, however, that this formula is not appropriate for neutrino energies below 1 GeV,  
where the cross-section energy dependence is $E^k$ with $k \ge 1$. This is, on the other hand, the typical range of energies 
considered for ``low'' $\gamma$ $\beta$-beams. The formula applies in the energy range considered in this paper, though.}.  
Secondly,  the ion half-life $T_{1/2}$ must  be long enough to accelerate the ions to the desired energy and short 
enough to allow a large number of them to decay in the storage ring such as to 
produce an intense neutrino beam. Eventually, assuming a limited space charge capacity of the storage ring, 
low-Z isotopes can be stored in larger number than high-Z isotopes \cite{Autin:2002ms}.

In Tab.~\ref{tab:ions} we remind the relevant parameters for four ions: $^{18}$Ne and $^6$He, $^8$Li and $^8$B.
\begin{table}
\begin{center}
\begin{tabular}{|c|c|c|c|c|} \hline \hline
   Element  & $A/Z$ & $T_{1/2}$ (s) & $Q_\beta$ eff (MeV) & Decay Fraction \\ 
\hline
  $^{18}$Ne &   1.8 &     1.67      &        3.41         &      92.1\%    \\
            &       &               &        2.37         &       7.7\%    \\
            &       &               &        1.71         &       0.2\%    \\ 
  $^{8}$B   &   1.6 &     0.77      &       13.92         &       100\%    \\
\hline
 $^{6}$He   &   3.0 &     0.81      &        3.51         &       100\%    \\ 
 $^{8}$Li   &   2.7 &     0.83      &       12.96         &       100\%    \\ 
\hline
\hline
\end{tabular}
\caption{\it $A/Z$, half-life and end-point energies for three $\beta^+$-emitters ($^{18}$Ne  and $^8$B)
and two $\beta^-$-emitters ($^6$He and $^8$Li). All different $\beta$-decay channels for $^{18}$Ne are presented~\cite{betadecays}.}
\end{center}
\label{tab:ions} 
\end{table}
As it was stressed in the literature (starting with Ref.~\cite{Zucchelli:sa}), $^6$He has the right half-life to be accelerated and stored
such as to produce an intense $\bar \nu_e$ beam using existing CERN infrastructures. 
According to the prescriptions given above, $^{18}$Ne has been identified as the best candidate as $\beta^+$-emitter, although its half-life
is twice that of $^6$He. Other ions were originally discarded for different reasons: for example,  $^{33}$Ar is too short-lived to be accelerated to 
the desired energy ($T_{1/2} = 0.17$ s).  As it can be seen in Tab.~\ref{tab:ions}, $^8$Li and $^8$B are good alternatives to 
$^6$He and $^{18}$Ne as $\beta^-$- and $\beta^+$-emitters, respectively. $^8$Li has similar half-life, $Z$ and $A/Z$ to $^6$He, 
thus sharing the key characteristics needed for the bunch manipulation. $^8$B has a lifetime similar to that of $^8$Li and $^6$He. 
Its  $A/Z$ is similar to that of $^{18}$Ne, instead, although its $Z$ is much smaller (which could in principle allow to store a larger 
amount of ions in the PS and in the SPS). This ion, however,  is difficult to produce with standard ISOLDE techniques (it reacts with many 
elements typically used in ISOL targets and ion sources and it is therefore barely released). 
For this reason it was originally discarded as a possible  $\beta^+$-emitter.
Notice that both $^8$Li and $^8$B have a much larger end-point energy than the two reference ions, though. As a consequence, at fixed $\gamma$, 
a longer baseline is needed to tune the $L/E$ ratio to the first oscillation peak with respect to $^6$He or $^{18}$Ne beams, and thus a smaller signal 
statistics is expected in the far detector. Therefore, the expected sensitivity to $\theta_{13}$ (at fixed flux and detector mass) of such beams is 
smaller than that for a ``standard'' beam with $\bar \nu_e$ and $\nu_e$ produced via $^6$He and $^{18}$Ne decays \cite{Donini:2006dx}.  
The end-point energy for $^8$Li and $^8$B decays being larger, nonetheless, a smaller $\gamma$ is needed to reach the desired neutrino energy 
in the laboratory  frame. Since the $\gamma$ choice depends in last instance on the facility that is used to accelerate the ions, 
it is then possible to reach higher neutrino energies using the same facility to accelerate the ions to be stored.

A detailed study of the attainable production rate of $^8$Li and $^8$B using ISOLDE techniques is lacking. Intense fluxes of both ions could  
be in principle produced using the ``ionization cooling'' technique proposed in Ref.~\cite{Rubbia:2006pi}, that will be studied 
in the framework of the EURO-$\nu$ Design Study, \cite{Euronu}. Notice that it is not difficult to produce an intense $^8$Li flux: 
using a thin Ta foil ISOL target, it is possible to produce $6 \times 10^8$ $^8$Li ions per $\mu$C, to be compared with $6 \times 10^6$ $^6$He ions per 
$\mu$C \cite{Tengblad}. The case of $^8$B is different: this ion was previously discarded as a $\beta^+$-emitter since it is extremely difficult 
to produce at a sufficient rate with ISOLDE techniques. However, using the ``ionization cooling'' technique, sustained $^8$Li and $^8$B production 
is supposed to be at reach through the reactions $^7$Li + D $\to$ $^8$Li + p and $^6$Li + $^3$He $\to$ $^8$B + n. 
We will assume in the rest of the paper that $^8$Li/$^8$B ion fluxes can be produced at least as efficiently
as $^6$He/$^{18}$Ne ones.

\subsection{Choice of $\gamma$}

Four classes of setups have been considered so far: $\gamma \simeq 10$ (``very low'' $\gamma$) \cite{Volpe:2003fi}, 
$\gamma \sim 100$ (``low'' $\gamma$), with a typical baseline of $O (100)$ Km for $^{18}$Ne and $^{6}$He 
\cite{Burguet-Castell:2003vv,Bouchez:2003fy,Donini:2004hu,Donini:2004iv,Campagne:2006yx} or 
$O (700)$ for $^{8}$B and $^{8}$Li \cite{Donini:2006dx,Rubbia:2006zv}; $\gamma \sim 300$ (``high'' $\gamma$),
with $L = O (700)$ Km \cite{Burguet-Castell:2003vv,Burguet-Castell:2005va,Donini:2006tt,Donini:2007qt,Huber:2005jk} 
for $^{18}$Ne and $^{6}$He or $O (7000)$ for $^{8}$B and $^{8}$Li \cite{Agarwalla:2006vf,Agarwalla:2007ai,Agarwalla:2006gz}; 
and $\gamma \geq 1000$ (very ``high'' $\gamma$), with baselines of several thousands kilometers, 
comparable with those suggested for the Neutrino Factory \cite{Burguet-Castell:2003vv,Huber:2005jk,Terranova:2004hu}. 
The three higher $\gamma$ ranges are related to different CERN-based facilities: the SPS, with $\gamma \leq 250$; the SPS+, 
with $\gamma \leq 600$; and the LHC, with $\gamma \geq 1000$. 

We recall that the SPS+ is a new synchrotron that uses fast cycling superconducting magnets, located in the SPS tunnel. 
Its construction has been proposed in the framework of the LHC maintenance and upgrade programme \cite{PAF}. 
Such a facility should be able to accelerate particles up to 1 TeV. 
Injecting protons into the LHC at 1~TeV strongly reduces the dynamic effects of persistent currents and stabilizes the operation of the collider. 
This would ease operation of the LHC and permit to increase luminosity
up to $10^{35}$~cm$^{-2}$s$^{-1}$ and, if needed, prepare it to double the operating energy (``DLHC'' phase) \cite{bruning}. 
Using the SPS+ as a final stage of acceleration for a $\beta$-Beam is not in conflict with LHC operations, since the SPS+ operates 
as injector only for a small fraction of its duty time (in the LHC filling phase).

The SPS+ as the final booster of the $\beta$-Beam is not the only possibility that can be envisaged to reach the multi-GeV regime. 
After injection of the ions from the SPS to the LHC, a mini-ramp of the LHC itself would bring the ions at $\gamma=350$. 
Differently from the previous case, this option would however require allocation of a significant fraction of the LHC duty cycle 
for neutrino physics, and it could be in conflict with ordinary collider operations. A third option is to use the TeVatron at FermiLab 
as the last acceleration stage (see, for example, Ref.~\cite{Jansson:2007nm}).

\subsection{Fluxes and cross-sections}

In the He/Ne $\beta$-Beams analyses it was shown that the minimal goal luminosities required for physics are 
$2.9 \times 10^{18}$ useful ion-decays per year for 
the $\beta^-$-emitter and $1.1 \times 10^{18}$ useful ion-decays per year for the $\beta^+$-emitter 
\cite{Burguet-Castell:2003vv,Burguet-Castell:2005va,Bouchez:2003fy}. 

Lacking a detailed study of the achievable $^8$Li and $^{8}$B fluxes, in the rest of the paper we consider three possible values 
for the $\beta$-beam flux\footnote{
Notice that, as in the NF-RS, we consider identical fluxes for the two detectors. As it is explained in App.~\ref{sec:storage}, 
this does not mean that the number of ions circulating in the storage ring must be twice the number considered for single detector
setups.}: 
\begin{itemize}
\item ``Nominal flux'' \\
In this case,  $2 \times 10^{18} $ decays per year per baseline for both $^8$Li and $^8$B are considered. 
These fluxes are close to the ``standard fluxes'', i.e. $2.9 \times 10^{18} $ and $1.1 \times 10^{18} $ decays per year for $^6$He and $^{18}$Ne.
\item ``Medium flux'' \\
In this case, fluxes of $5 \times 10^{18} $ decays per year per baseline for both ions are assumed. 
\item ``Ultimate flux'' \\
For this most optimistic scenario, fluxes of $10 \times 10^{18} $ decays per year per baseline for both ions are considered.
\end{itemize}

We remind that an intense $^8$Li (but not $^8$B) beam could be produced using well-studied ISOLDE techniques. An interesting option 
could be to start a first phase with an intense $\bar \nu_e$ flux aiming at the two detectors for a five year period, 
whilst building the facility needed to produce the $^8$B beam that would be used in the second phase.

As it is shown in the Appendix, no specific drawback in the acceleration and storage phase for using Li/B instead 
of the standard He/Ne ions is expected. For this reason, we assume that the ``nominal flux'' could be safely 
used at this setup, if shown to be achievable for standard setups. An increase of the ion flux up to the ``ultimate flux'' is believed to be possible (see Ref.~\cite{Lindroos2}). Notice, moreover, that due to the higher energy of this setup compared to standard He/Ne options, the atmospheric neutrino background
is expected to be significantly lower and a larger number of bunches can be thus injected into the storage ring, as it will 
be explained at the end of this section.

\begin{figure}[t!]
\vspace{-0.5cm}
\begin{tabular}{cc}
\hspace{-0.55cm} \epsfxsize7.5cm\epsffile{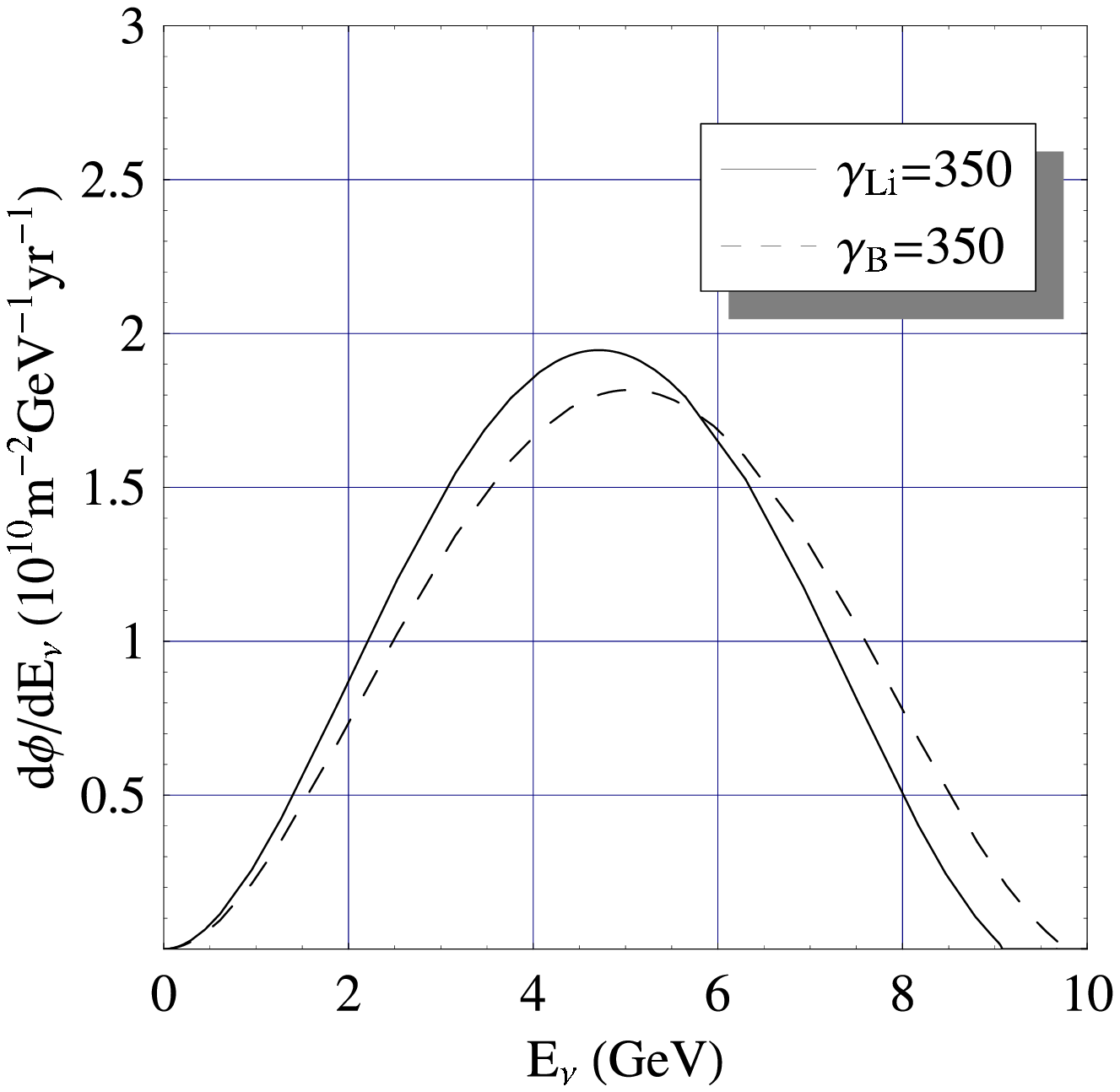} & 
                 \epsfxsize7.5cm\epsffile{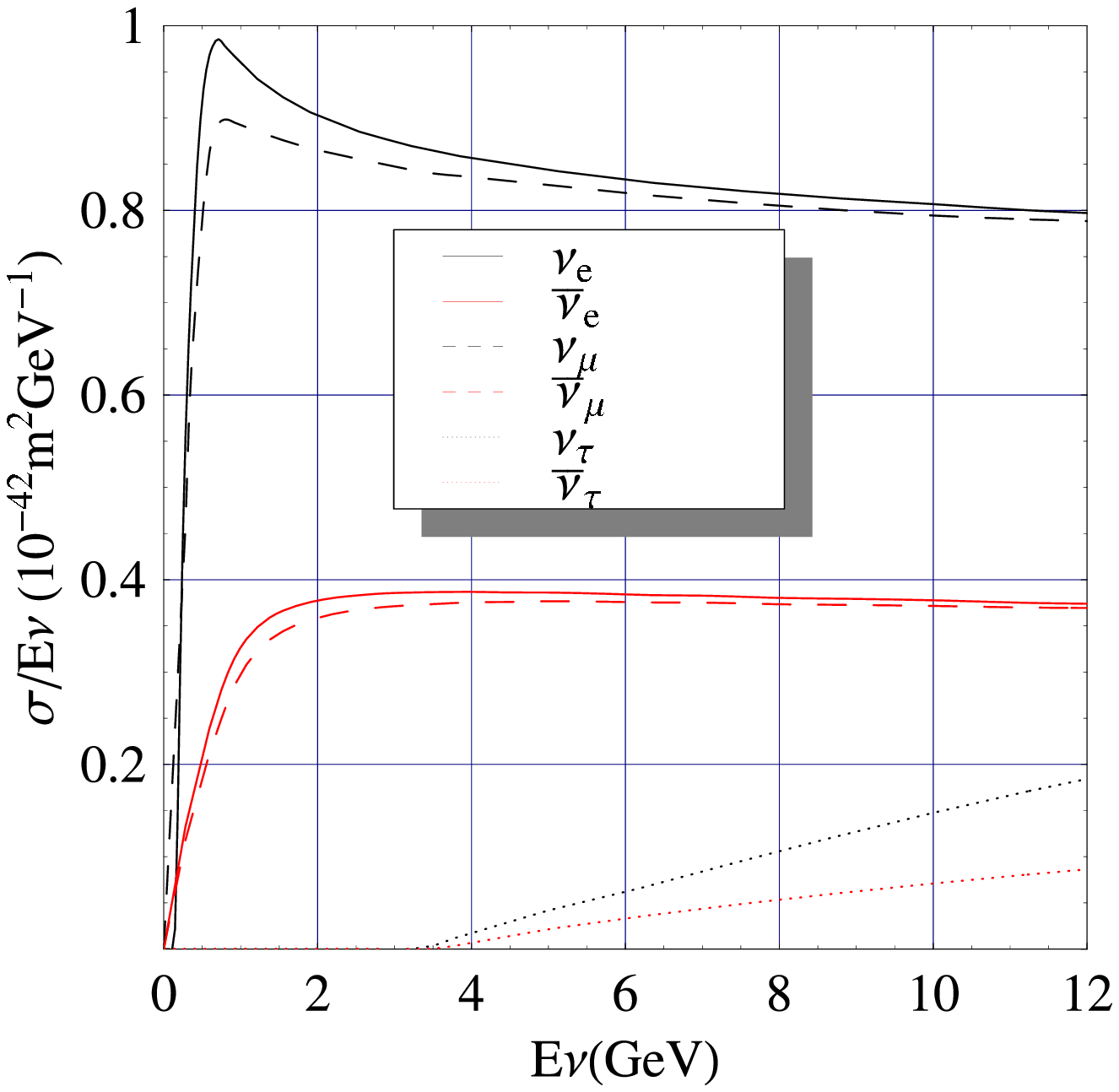}
\end{tabular}
\caption{\it 
Left: $\beta$-beam fluxes at a $L = 2000$ Km baseline as a function of
the neutrino energy for $^8$B (dashed line) and $^8$Li (solid line), for an integrated flux of $10 \times 10^{18}$ useful decays per year. 
Right: the $ \nu N$ and $\bar \nu N$ charge current total cross-sections on iron, \cite{lipari}.}
\label{fig:fluxes}
\end{figure}

We have considered identical fluxes for $^8$Li and $^8$B ions, in the absence of a clear indication of a significant
asymmetry in the ion production stage (differently from the He/Ne case, see Appendix). 
Neutrino fluxes for the Li/B $\beta$-beam computed at $L = 2000$ Km from the source for the ``ultimate flux'' are shown in Fig.~\ref{fig:fluxes}(left).

In Fig.~\ref{fig:fluxes}(right) the $\nu$ and $\bar \nu$ cross-sections on iron used, taken from Ref.~\cite{lipari}, are shown. 
Notice that, for the setup proposed in this paper, most of the neutrinos have multi-GeV energies, and therefore the $\nu$N cross-sections are
dominated by the deep inelastic part. Thus, the details of the different cross-sections present in the literature are not as relevant as it was
in the case for lower energies neutrino beams. 

\subsection{The detector}

Traditional technologies for $\nu$ production (conventional beams and superbeams) allow the investigation of the 1-3 sector 
of the leptonic mixing matrix through the appearance of $\nu_e$ and $\bar{\nu}_e$ at baselines $\ge 100$~km, i.e. through 
the information encoded in the $\nu_\mu \rightarrow \nu_e$ and $\bar{\nu}_\mu \rightarrow \bar{\nu}_e$ transitions probabilities. 
In that context, optimal far detectors are low-density, massive electromagnetic calorimeters
(liquid scintillators, water \v Cerenkov or liquid Argon TPC's~\cite{nostrareview}). 

On the other hand, both $\beta$-Beams and Neutrino Factories exploit the T-conjugate channel 
$\nu_e \rightarrow \nu_\mu$ and $\bar{\nu}_e \rightarrow \bar{\nu}_\mu$.  
An important difference of the $\beta$-Beam with respect to the Neutrino Factory is that in the former only 
$\nu_e (\bar \nu_e)$ are present in the beam, whereas in the latter both $\nu_e$ and $\bar \nu_\mu$ (or $\bar \nu_e, \nu_\mu$) are present. 
Therefore, in a $\beta$-Beam--based experiment final lepton charge identification is not needed. 
This usually allows for the use of large water \v Cerenkov detectors, something impossible at the Neutrino Factory, 
where magnetized detectors are mandatory when looking for $\nu_e \to \nu_\mu$ oscillations. 
In both cases, calorimetric measurements are needed to reconstruct the
neutrino energy\footnote{The only notable exception concerns ``monochromatic Beta Beams''\cite{monocromatic} based on ions decaying
through electron capture. Energy reconstruction with these beams serves to suppress backgrounds, though.}. 
In our setup neutrinos are produced by the $\beta$-Beam with a (relatively) high energy ($E_\nu \sim 6$ GeV)
with respect to the low-$\gamma$ ($E_\nu \sim 300$ MeV) and the high-$\gamma$ ($E_\nu \sim 1$ GeV) $^6$He/$^{18}$Ne setups.
For the high-$\gamma$ $^8$Li/$^8$B $\beta$-Beam, the use of dense detectors is therefore possible. 
In particular, the choice of the passive material of the calorimeter depends on the typical range of the
primary muon; the latter must be significantly larger than the interaction length to allow for filtering 
of the hadronic part and effective NC and $\nu_e$~CC selection. 
For neutrinos of energies greater than $\sim$1~GeV, iron offers the desired properties. 
As a consequence, the energy reached at the SPS+ can be exploited to switch from a low-Z to a high-Z/high-density 
calorimeter also in the case of the $\beta$-Beam. The use of iron detectors avoids the need for large
underground excavations, which are mandatory for $\beta$-Beams of lower $\nu$ energies. 
Since these detectors are capable of calorimetric measurements, they can be exploited even better 
than water \v Cerenkov to obtain spectral informations. They are not expected to reach, anyhow,
the granularity of liquid argon TPC's or the megaton-scale mass of water \v Cerenkov's. 
Hence, in spite of the underground location, they cannot be used for proton decay measurements and low-energy
astroparticle physics.

Several techniques can be employed for the design of the active detectors of large mass iron calorimeters. 
A detailed study of a magnetized iron detector suitable for a $\gamma = 350$ $^6$He/$^{18}$Ne $\beta$-Beam was
performed in Ref.~\cite{Donini:2006tt}. The design of the detector was based on glass
Resistive Plate Chambers (RPC). In the context of the $\beta$-Beam, the advantage of using RPC's 
mainly reside on their low production cost, along the line investigated by the MONOLITH~\cite{monolith} 
and INO~\cite{ino} collaborations. 
Magnetization of the detector could in principle be removed in the present analysis, since at the $\beta$-Beam 
no significant $\bar \nu_\mu$ flux that must be distinguished from the $\nu_e \to \nu_\mu$ signal
 is produced. Removing the magnetization of the detector can be used to reduce
costs on the detector side, if it is not necessary to reduce possible backgrounds. 

\subsection{Signal event rates}

In Tab.~\ref{tab:rates} we show the total expected event rates
when each detector (assuming perfect efficiency) is exposed to the ``ultimate flux'' for five years of data taking, for several choices of $\sin^22 \theta_{13} $, $\delta$ and the mass hierarchy. 
The other oscillation parameters are fixed to the following values: $\theta_{12}=33^\circ$, 
$\theta_{23}=45^\circ$, $\Delta m^2_{21}=8\times10^{-5}$ eV$^2$ and $\Delta m^2_{31}=2.6\times10^{-3}$ eV$^2$. The non-oscillated expected rates are also shown. 
Notice that we do not study $\nu_e$ disappearance data in our setup (see Ref.~\cite{Agarwalla:2006gz} for such a study at the L = 7000 Km baseline) . 

\begin{table}[hbtp]
\begin{center}
\begin{tabular}{|c|c|c|c|c|c|c|c|} \hline
  $\sin^22 \theta_{13} $ & $\delta$ &  $s_{atm}$ & $N_{\nu_\mu}^{2000}$ &
  $N_{\bar{\nu}_\mu}^{2000}$ & $N_{\nu_\mu}^{7000}$ &
  $N_{\bar{\nu}_\mu}^{7000}$  \\ \hline \hline
  $0.01$ & $90^\circ$ & + & 595 &  20 & 155 & 2\\ \hline
  $0.01$ & $-90^\circ$ & + & 206 &  103 & 170 & 1\\ \hline
  $0.01$ & $90^\circ$ & - & 235 &  95 & 3 & 67\\ \hline
  $0.01$ & $-90^\circ$ & - & 47 &  274 & 6 & 62\\ \hline 
  $0.001$ & $90^\circ$ & + & 139 &  10 & 15 & 0\\ \hline
  $0.001$ & $-90^\circ$ & + & 15 &  36 & 20 & 0\\ \hline
  $0.001$ & $90^\circ$ & - & 81 &  7 & 0 & 7\\ \hline
  $0.001$ & $-90^\circ$ & - & 21 &  64 & 2 & 6\\ \hline \hline
$\nu_e (\bar \nu_e)$ non-osc/$10^3$& & & 61.4 & 26.2 &  5.0 & 2.1 \\ \hline \hline
\end{tabular}
\end{center}
\caption{\label{tab:rates} \it Event rates (with perfect efficiency) for a 5 years exposure to a $10 \times 10^{18}$ ion decays per year flux, 
for several choices of $\sin^22 \theta_{13} $, $\delta$ and the mass hierarchy.  The expected number of non-oscillated events at the two 
baselines are also given.
}
\end{table}

Notice the strong complementarity between the two baselines:
\begin{itemize}
\item
The event rates at the 2000 Km baseline show a strong dependence on both the CP-violating phase $\delta$ and the sign of the atmospheric mass difference. 
Neutrino (antineutrino) events are enhanced for positive (negative) values of $\delta$ and normal (inverted) hierarchy. 
This strong dependence on both unknowns is also the source of strong degeneracies, when the effect of $\delta$ is able to compensate that of 
the mass hierarchy. Such a situation can be seen in the second and third rows of Tab.~\ref{tab:rates}, where very similar event rates are found 
at the $L = 2000$ Km baseline for $\delta = -90^\circ$ and $s_{atm} =  +$ and for $\delta = 90^\circ$ and $s_{atm} = -$. 
For this reason, with only one detector the sensitivity to the mass hierarchy would be limited to positive (negative) values of $\delta$ 
for a normal (inverted) hierarchy, where the effects of  $\delta$ and the mass hierarchy push in the same direction (see eg. Fig.~(19) of Ref.~\cite{Donini:2005db}). 
\item
On the other hand, it can be seen that the event rates at the $L = 7000$ Km baseline are practically insensitive to the CP-violating phase $\delta$ 
(as this baseline is so close to the ``magic baseline'', where $\delta$-dependence vanishes). 
The dependence on the mass hierarchy is, however, very strong, with a nearly resonant enhancement of the neutrino (antineutrino) 
oscillation probability if the hierarchy is normal (inverted), regardless of the value of $\delta$ 
(see Refs.\cite{Agarwalla:2006vf,Agarwalla:2007ai,Agarwalla:2006gz}). 
\end{itemize}
The combination of both baselines can thus provide an unambiguous determination of both the CP violation phase and the mass hierarchy 
for large regions of the parameter space.

\subsection{Backgrounds, efficiencies and systematic errors}

The $\beta$-Beam flux illuminating the detector can be considered, with a very high accuracy, a pure $\nu_e$ beam. 
An undesired $\nu_\mu$ and $\bar \nu_\mu$ beam contamination could in principle originate from the daughter ions produced
in the $\beta$-decay that collide with the storage ring magnets, acting as a fixed target. 
This background was studied in Ref.~\cite{Zucchelli:sa} for $^6$Li ions and it is smaller than $10^{-4}$. 
The beam contamination induced by $^{18}$F and $^8$Be ions has not been studied in detail, but it is supposed
to be similar to that of $^6$Li, and thus negligible. 

When looking for $\nu_e \to \nu_\mu$ oscillations at a $\beta$-Beam, the main source of beam background are $\nu_e$ CC interactions 
(with a non-observed electron) or NC interactions in which a pion or some other meson produced in the hadronic shower mimics a muon track. 
Another, sub-dominant, possible background source are $\nu_e$ CC (again, with a non-observed electron) 
or NC interactions in which a charmed-meson is produced that eventually generates a muon through a semileptonic decay. 
It is clear that measuring the charge of the muon will strongly reduce both backgrounds (for $\nu_e N \to e^- D^+ X$, the final $\mu^+$ has 
opposite charge with respect to the signal, $\nu_\mu N \to \mu^- X$). 

A full simulation of the response of a magnetized iron detector to the beam proposed in this paper is lacking. In Ref.~\cite{Donini:2006tt} the signal identification efficiency of such a detector for $^6$He boosted at $\gamma= 350$ and $^{18}$Ne boosted at $\gamma= 580$ (i.e. for a neutrino energy around 1 GeV) was found to be of the order of 50-60\%. 
On the other hand,  in the framework of the ISS report \cite{Group:2007kx}, a detailed
study of the MIND detector exposed to the Neutrino Factory beam (i.e. for a neutrino energy around 30 GeV) has been presented, finding  a $\nu_\mu$ identification efficiency in the energy range of interest 
as high as 70\%.



In Ref.~\cite{Donini:2006tt}, it was found that the probability for the background to mimic a CC-like event is around 1\%.  A rather large beam-induced background was therefore expected for this setup, as the consequence of the limited pion rejection capability of this detector compared with more challenging $\beta$-Beam or Neutrino Factory detector designs. This large background was mainly caused by the (relatively) low energy of the neutrinos. At the typical neutrino energy of a  
$^8$Li/$^{8}$B $\gamma=350$ $\beta$-Beam, these backgrounds are much easier to suppress 
in iron calorimeters.  Consistently, in Ref.~\cite{Agarwalla:2006vf} this background was completely neglected, on the basis of the simulations by the INO collaboration. 
Moreover, in Refs.~\cite{Cervera:2000vy,Cervera:2000kp,cerveragolden07} the fractional backgrounds for a 50 GeV Neutrino Factory beam targeting 
an iron calorimeter were found to be around or below $10^{-4}$ for the region around 5 GeV. Since in our setup there is no such a strong down-feed of the background from high energy neutrinos, we expect $10^{-4}$ to be a pessimistic upper limit for the beam-induced background.

In the numerical analysis below, event rates have been divided into nine bins between 1.5 and 10.5 GeV, with $\Delta E = 1$ GeV. The detector energy resolution has been implemented through a gaussian resolution function with $\sigma = 0.15 \times E$.  We have considered a constant $\nu_\mu/\bar \nu_\mu$ identification efficiency of $65\%$  and a constant fractional background equal to $10^{-5}$ of the unoscillated events per bin. We have also studied the impact of the 
beam background on the physics performance of the setup increasing the fractional 
background up to $10^{-4}$,  showing explicitly that the effect is small for any of the considered observables. 

A separate discussion is required for the atmospheric neutrinos that interact inside the detector or in its proximity, giving rise to muon events that can be confused with the signal. 
This background was studied at a $\beta$-Beam, both for water \v Cerenkov \cite{Burguet-Castell:2003vv,Mezzetto:2003ub} 
and iron detectors \cite{Jansson:2007nm,Donini:2006tt,Donini:2007qt}. The number of muons produced by atmospheric neutrinos 
crossing the detector aligned with the $\beta$-Beam flux was found to be of the order of tens of events per Kton per year.
This background would completely dominate the oscillation signal (see Tab.~\ref{tab:rates}). It is therefore mandatory to reduce it by a proper timing of the ion bunches. 

In order to have a good time correlation of the signal with the neutrino flux produced at the source, the ions circulating in the storage ring
must occupy a small fraction of it. The "suppression factor" ($S_f$) is defined as:
\begin{equation}
S_f  = \frac{v \times \Delta t_{bunch} \times N_{bunch}}{L_{ring}}
\end{equation}
where $v\simeq c$ is the ion velocity. For $^6$He/$^{18}$Ne ions boosted at $\gamma= 100$ ($E \sim 300$ MeV),
the suppression factor must be $S_f \sim 10^{-3}$. Such a tight $S_f$ can be achieved with a
challenging $\Delta t_{bunch}$=10~ns time-length, with a maximum of $N_{bunch}=8$ 
bunches circulating at the same time. 

At higher energies (e.g. $\gamma=350$ for $^6$He, $E \sim 1.2$ GeV), however, the atmospheric background is suppressed by about 
one order of magnitude with respect to $\gamma=100$. $S_f$ can thus be correspondingly relaxed.\footnote{Notice that the suppression factor scales with $1/\gamma$.
Thus, in the case of ions circulating at $\gamma = 350$, the bunch-length 
$(v \times \Delta t_{bunch})$ gets Lorentz-contracted. Therefore, maintaining the same $\Delta t_{bunch}$ 
as for the He/Ne setups, we can inject a larger number of bunches into the storage ring, thus increasing the neutrino flux 
(provided that the injection system can sustain the increased request of bunches and/or ions per bunch).}
The average neutrino energy for high-$\gamma$ $^8$Li/$^8$B ion beams being $E \sim 6$ GeV,
moreover, an extra suppression of the atmospheric neutrino background is expected
with respect to the  high-$\gamma$ $^6$He/$^{18}$Ne ion beams.
As it can be seen from Fig.~2 of Ref.~\cite{Ashie:2005ik}, for example, the atmospheric 
neutrino flux decreases about two orders of magnitude passing from 1 to 6 GeV. 
As a consequence, less demanding bunch time-lengths are acceptable for the setup proposed in this paper, 
thus simplifying the storage ring design. For this reason, in the numerical analysis presented in Section~\ref{sec:results} we have neglected 
the atmospheric neutrino induced background. 

We have, eventually,  considered a 2.5\% and 5\% systematic errors on the signal and on the beam-induced background, respectively. 
They have been included as ``pulls'' in the statistical $\chi^2$ analysis. The effect of increasing these errors to 10\% and 20\%, 
respectively, was also considered. 
It has been found that the impact is negligible. 

The following $1 \sigma$ errors for the oscillation parameters were also considered: $\delta \theta_{12} = 1\%$, $\delta \theta_{23} = 5\%$, $\delta \Delta m^2_{21} = 1 \%$ and $\Delta m^2_{31} = 2\%$. 
If $\theta_{23}$ turns out to be maximal the error on $\theta_{23}$ could be larger than the $5\%$ we assumed.
We studied the effect of increasing the error to $10\%$, which is almost the present uncertainty, given that this parameter
will not be measured by the proposed setup. We have checked that our results are not significantly affected by considering such an error. 
Eventually, an error $\delta A = 5\%$ has been considered for the Earth density given by the PREM model. Marginalization over these parameters has been performed for all observables. The Globes 3.0 \cite{Globes} software was used to perform the numerical analysis.

\section{Results} 
\label{sec:results}

In this section we present the performance of the considered setup on some of the items in the ``shopping list''. Not all of them are accessible at this setup. In particular, to probe possible deviations of $\theta_{23}$ from maximal mixing and eventually measure the $\theta_{23}$-octant, very precise measurements of $\theta_{23}$ and $\Delta m^2_{atm}$ are required. 
These are best achieved through the $\nu_\mu \rightarrow \nu_\mu$ and $\nu_\mu \rightarrow \nu_\tau$ channels. These channels are not accessible at $\beta$-Beams, though, since  only $\nu_e$ are present in the beam. We will thus not address these measurements in this section. 
These oscillations however occur in atmospheric neutrinos and could be studied in the two detectors
proposed in this setup. The combination of this data with the study of the golden channel of the $\beta$-Beam could provide some sensitivity to these unknowns (see, e.g., Refs.~\cite{Campagne:2006yx,Choubey:2005zy,Huber:2005ep}). 

On the other hand, the golden channel has excellent sensitivity to $\theta_{13}$, $\delta$ and the mass hierarchy.
Lacking a detailed study of the maximum achievable $^8$Li and $^{8}$B fluxes, we will present the results 
for three fluxes, as defined in Sect.~\ref{sec:det}.
As we will see, this is the key factor limiting the sensitivity of the setup, since its very long baselines limit 
the statistics at the detectors. We have also studied the impact of  the beam-induced background 
and of the systematic errors on the performance of the experiment. We will show that these uncertainties 
do not affect significantly the physics reach of the setup.

\subsection{$\theta_{13}$ discovery potential}

\begin{figure}[t!]
\vspace{-0.5cm}
\begin{center}
\begin{tabular}{cc} 
\hspace{-0.55cm} \epsfxsize7.5cm\epsffile{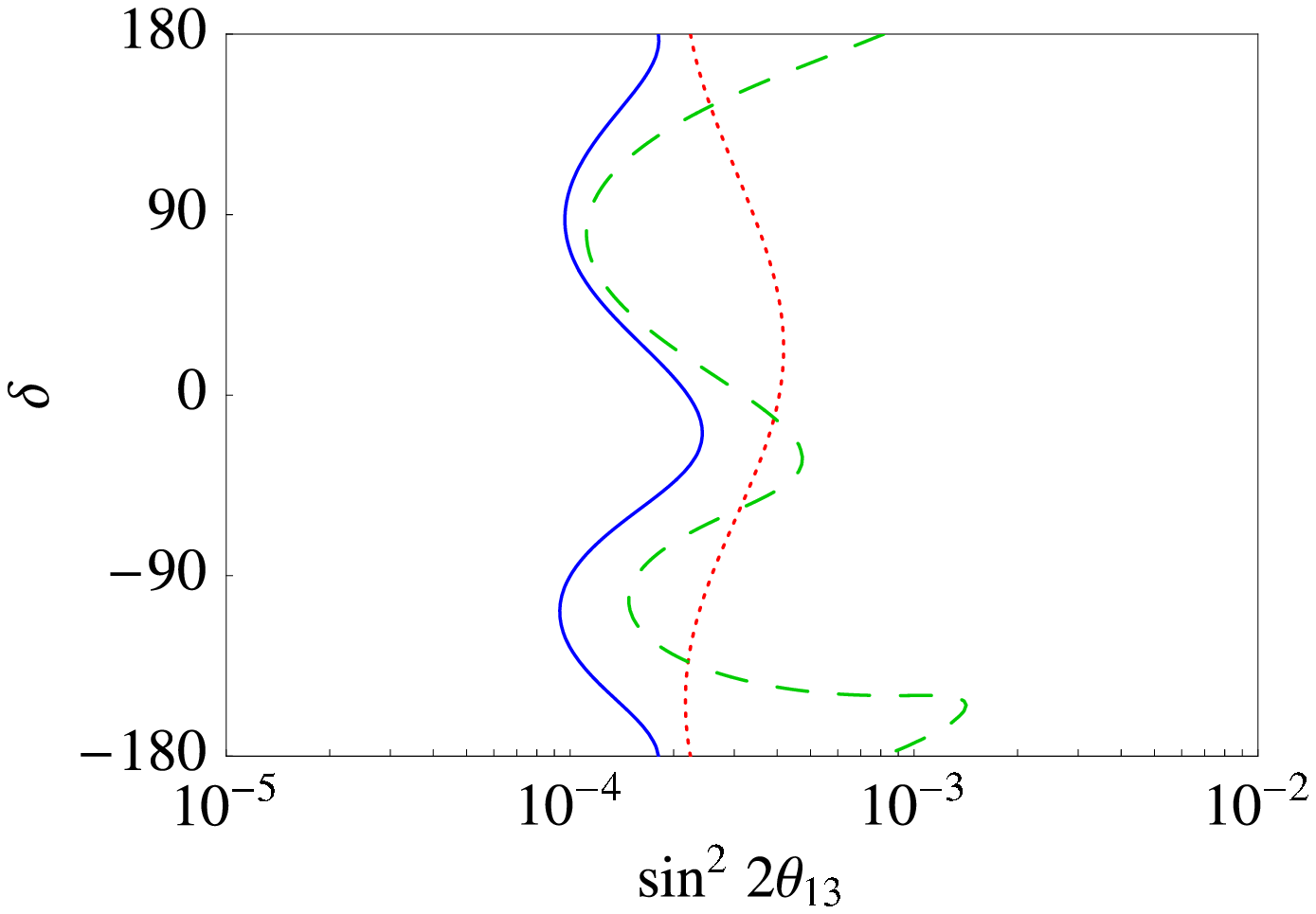} & 
                 \epsfxsize7.5cm\epsffile{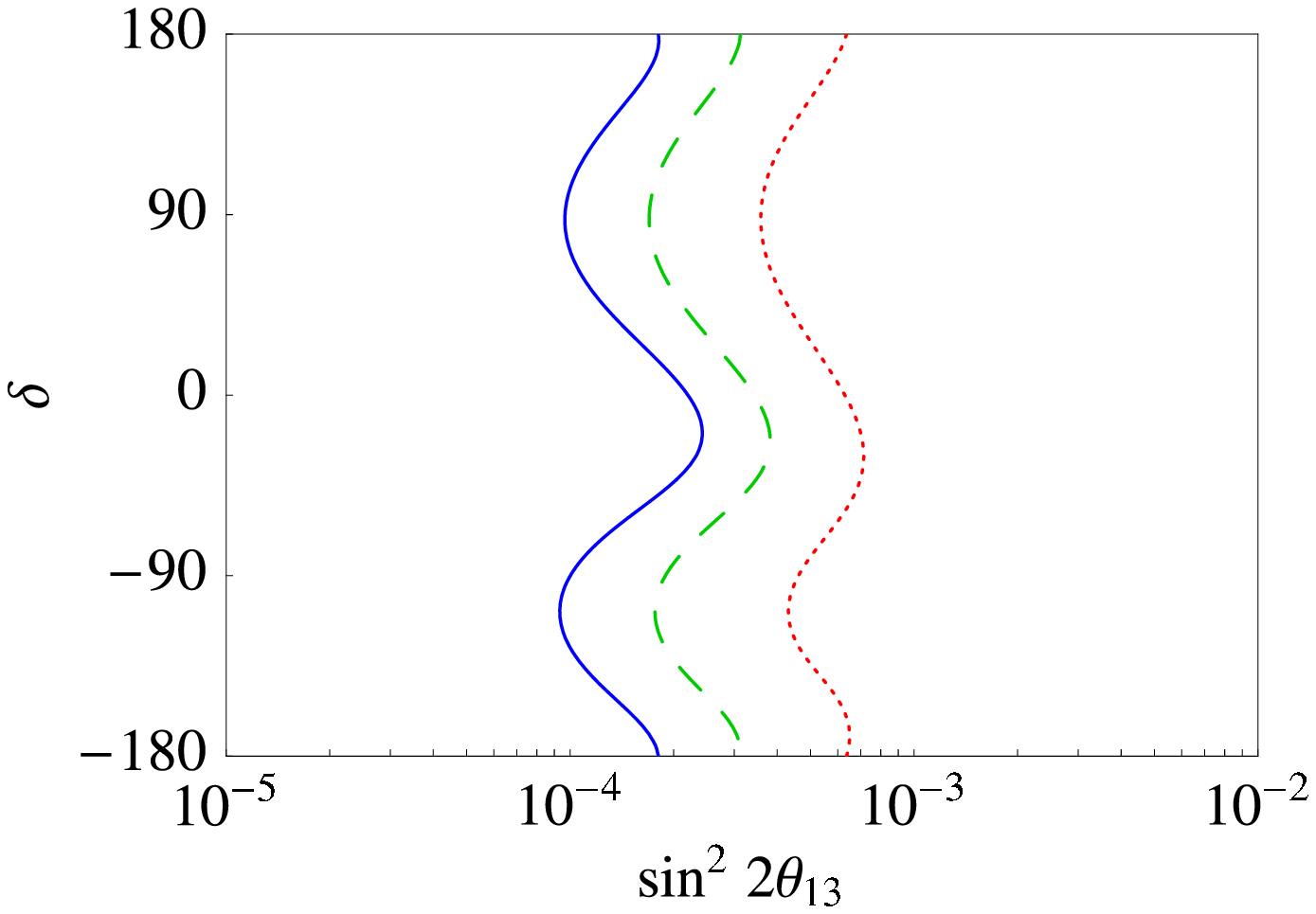} \\
\hspace{-0.55cm} \epsfxsize7.5cm\epsffile{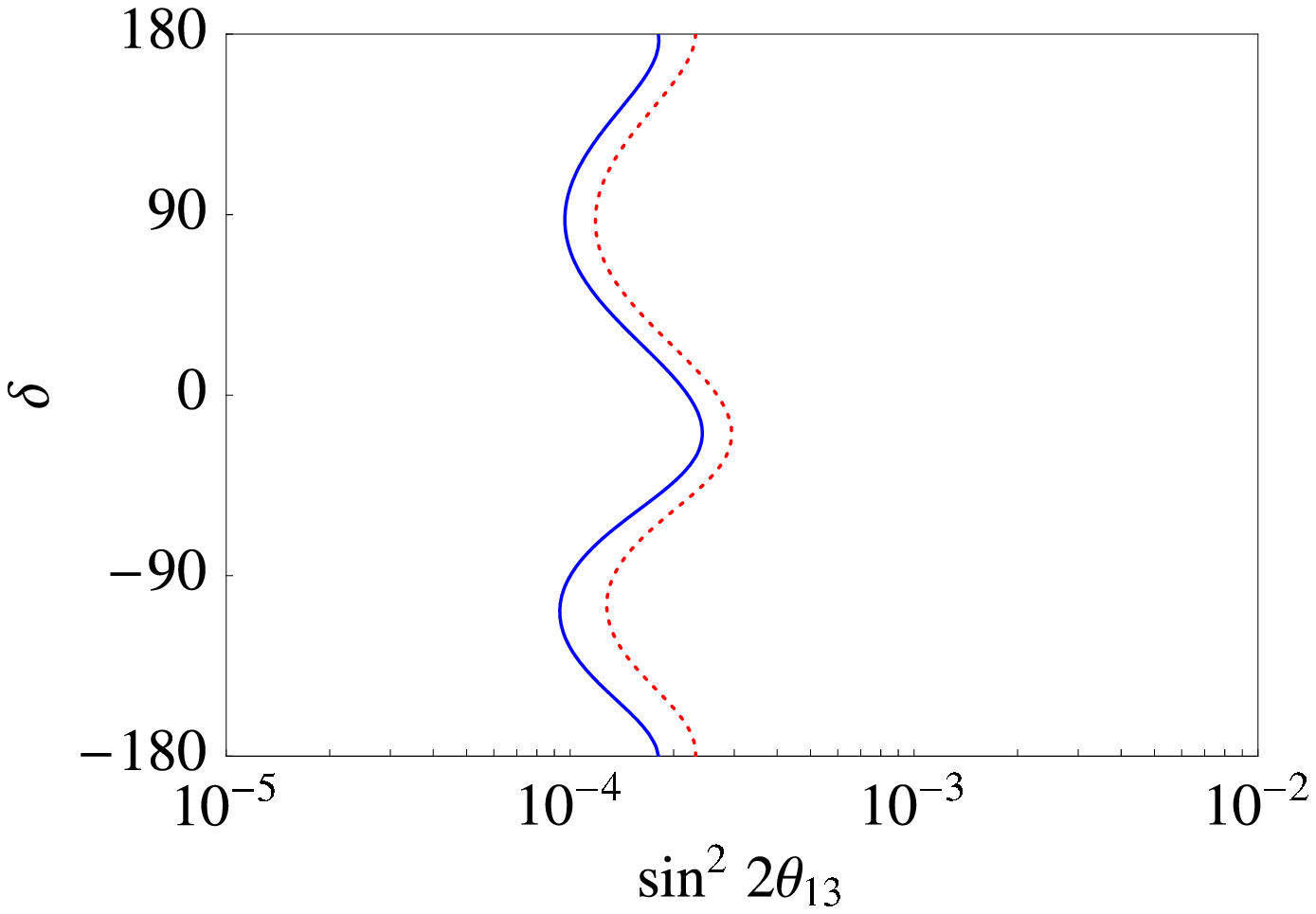} & 
                 \epsfxsize7.5cm\epsffile{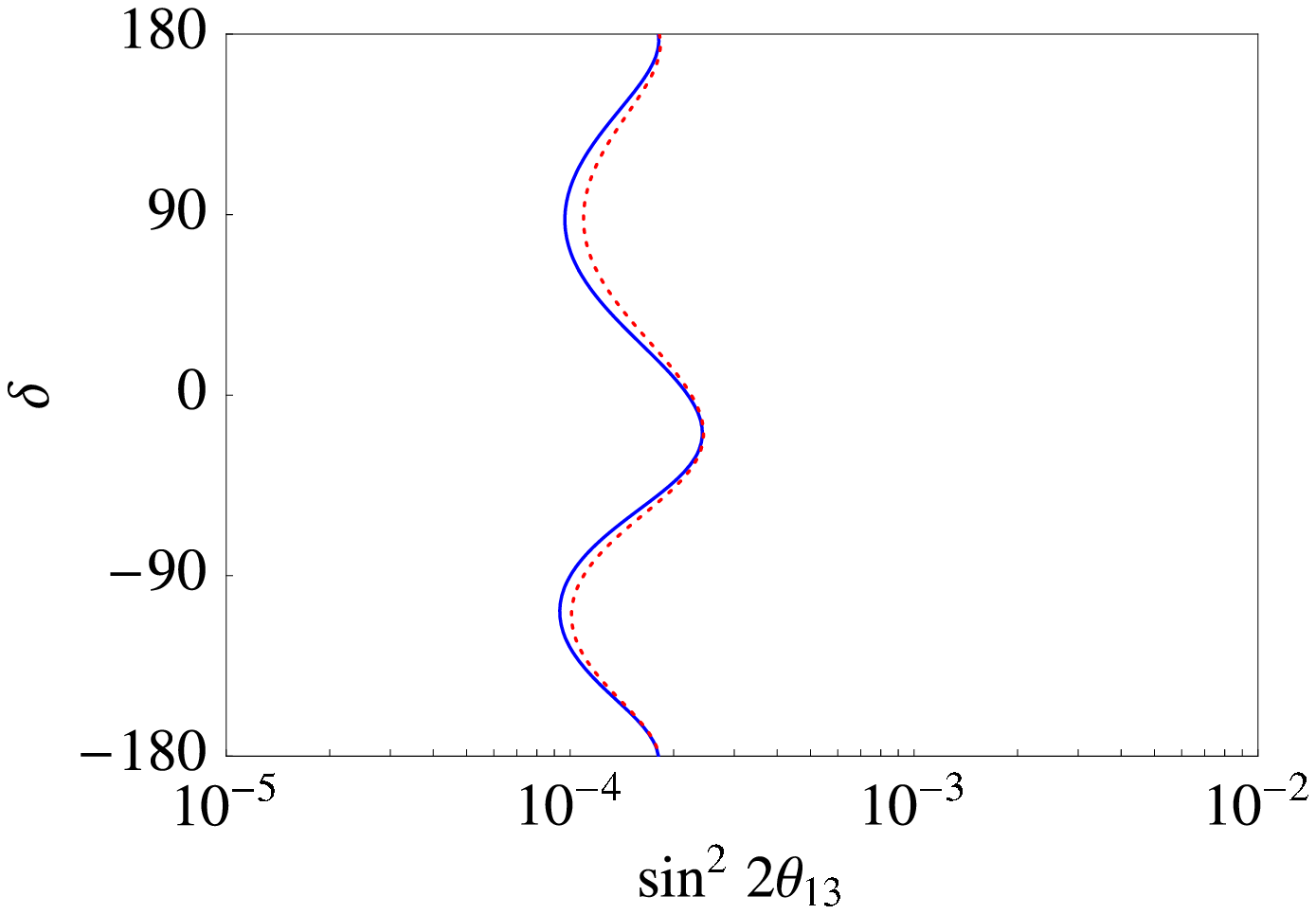}
\end{tabular}
\caption{\it 
3$\sigma$ $\theta_{13}$ discovery potential. 
Top left: comparison of baselines for a flux of  $1 \times 10^{19}$ useful ion decays per year. 
Dashed stands for $L = 2000$ Km; dotted for $L = 7000$ Km and solid for the combination of the two;
Top right: the impact of the flux for the combination of the two baselines.
Dotted stands for $2 \times 10^{18}$, dashed for $5 \times 10^{18}$ and solid 
for $1  \times 10^{19}$ useful ion decays per year;
Bottom left: the impact of the beam-induced background for the combination of the two baselines
and a flux of $1 \times 10^{19}$ useful ion decays per year.
Dotted stands for a background of  $10^{-4}$ times the non-oscillated events, solid for
$10^{-5}$ times the non-oscillated events;
Bottom right: the impact of systematic errors for the combination of the two baselines and a flux of $1 \times 10^{19}$ useful ion decays per year. 
Solid stands for systematics of 2.5\% and 5\% on 
the signal and the background, respectively: dotted for systematics of 10\% and 20\% on 
the signal and the background, respectively.}
\label{fig:t13sens}
\end{center}
\end{figure}

In Fig.~\ref{fig:t13sens} we present our results for the $\theta_{13}$ discovery potential, defined in the following way: the values of $\theta_{13}$ and $\delta$ in the plots represent the ``true'' values of these parameters, i. e. the input values assumed to generate the number of events that would be measured at the detector. A ``true'' normal hierarchy is also assumed. For each of these input values, the $\chi^2$ for $\theta_{13}=0$ (marginalized in the rest of the parameters) was computed. 
If the value of the $\chi^2 > 9$,  then the hypothesis that 
$\theta_{13}=0$ can be rejected at 3$\sigma$ for those ``true'' values of $\theta_{13}$ and $\delta$.

In the top left  panel we present the sensitivities to $\theta_{13}$ of the two baselines considered. 
The (green) dashed line corresponds to the sensitivity to $\theta_{13}$  with the detector at 2000 Km.
The maximal sensitivity, $\sin^22\theta_{13} \geq 1.5\times10^{-4}$, is achieved for 
$\delta = 90^\circ$ and $\delta = -90^\circ$, when event rates for neutrinos and antineutrinos peak, respectively.  
The (red) dotted line is  the sensitivity to $\theta_{13}$  with the detector at 7000 Km.
Notice that, in spite of the longer baseline, the sensitivity is similar to the one achievable with 
the 2000 Km detector. This can be understood from the resonant enhancement of the mixing angle through matter effects at this baseline. 
Notice also that the $\delta$ dependence of the sensitivity is much milder, since the detector is located near the magic baseline, 
where the terms involving $\delta$ vanish. 
Eventually,  the (blue) solid curve is the sensitivity to $\theta_{13}$ for the combination of the two baselines. In this case,  
$\theta_{13}$ can be measured for any value of $\delta$ provided that 
$\sin^22\theta_{13} > 2\times10^{-4}$. 

In the top right panel we study the dependence of the $\theta_{13}$-sensitivity on the neutrino flux. 
Fluxes of $2 \times 10^{18}$ (red dotted line), $5 \times 10^{18}$ (green dashed line) and $1 \times 10^{19}$ (blue solid line) 
useful ion-decays per year per baseline have been considered, for the combination of the two baselines.  
The more or less linear increase of the sensitivity with the flux indicates that the experiment performance is statistics-dominated.

The bottom left panel shows the impact  of the beam-induced background on the $\theta_{13}$-sensitivity for the combination of the 
two baselines for a flux of $1 \times 10^{19}$ useful ion decays
per year. Backgrounds of $10^{-5}$ (blue solid line) and $10^{-4}$ (red dotted line) of the total unoscillated events are considered. 
Notice that even increasing the background by an order of magnitude the loss of sensitivity is very small. On the other hand, decreasing 
the fractional background below $10^{-5}$
has no effect whatsoever. This background is thus equivalent in practice to no background.

Eventually, in the bottom right panel we present the impact of the systematic errors  on the $\theta_{13}$-sensitivity for the 
combination of the two baselines for a flux of $1 \times 10^{19}$ useful ion decays
per year. The systematic errors are increased from 2.5\% and 5\% (blue solid line) on the signal 
and the background, respectively, to 10\% and 20\% (red dotted line). It can be seen that the impact
of systematic errors is negligible. 

\subsection{CP discovery potential}

\begin{figure}[t!]
\vspace{-0.5cm}
\begin{center}
\begin{tabular}{cc} 
\hspace{-0.55cm} \epsfxsize7.5cm\epsffile{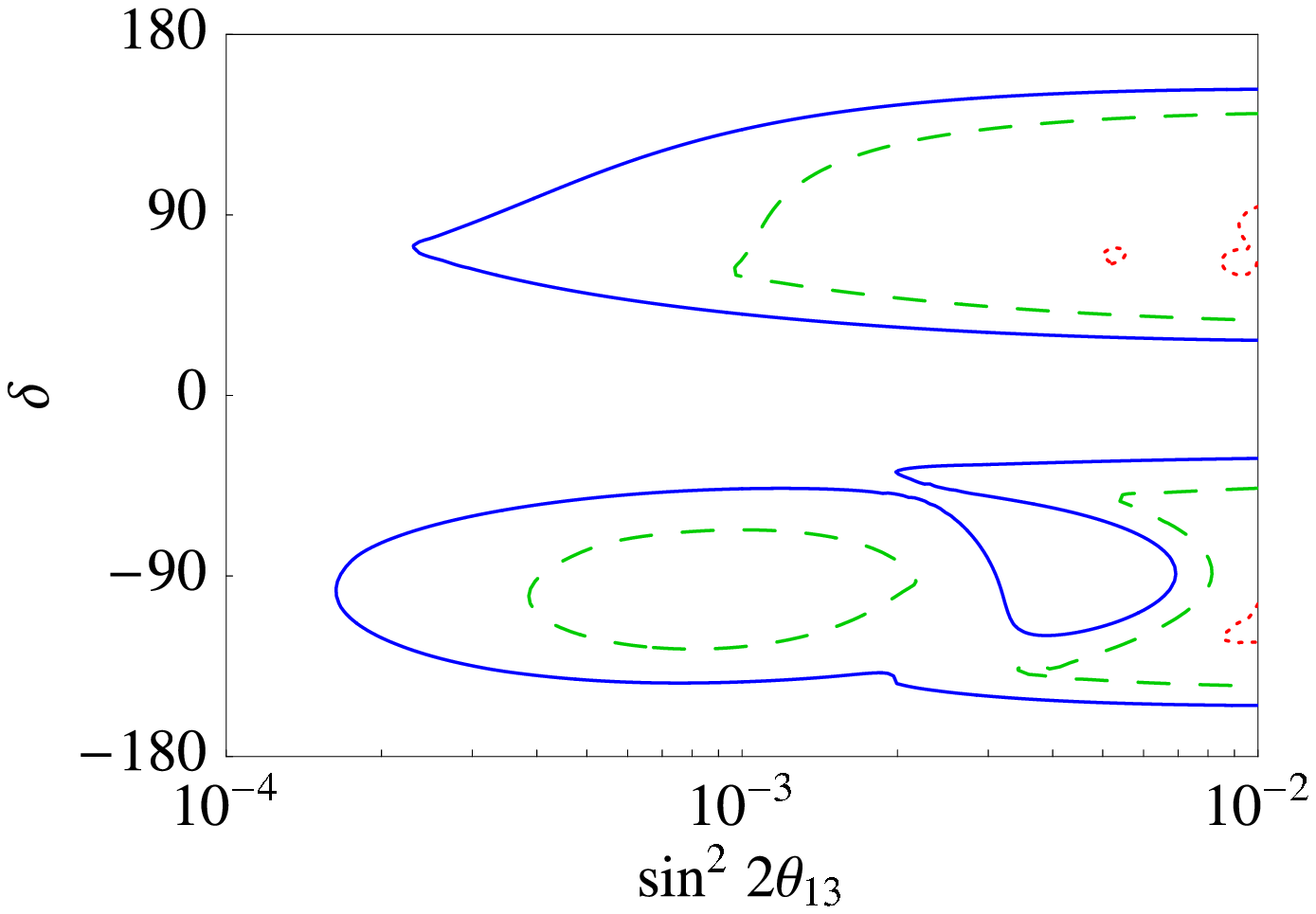} & 
                 \epsfxsize7.5cm\epsffile{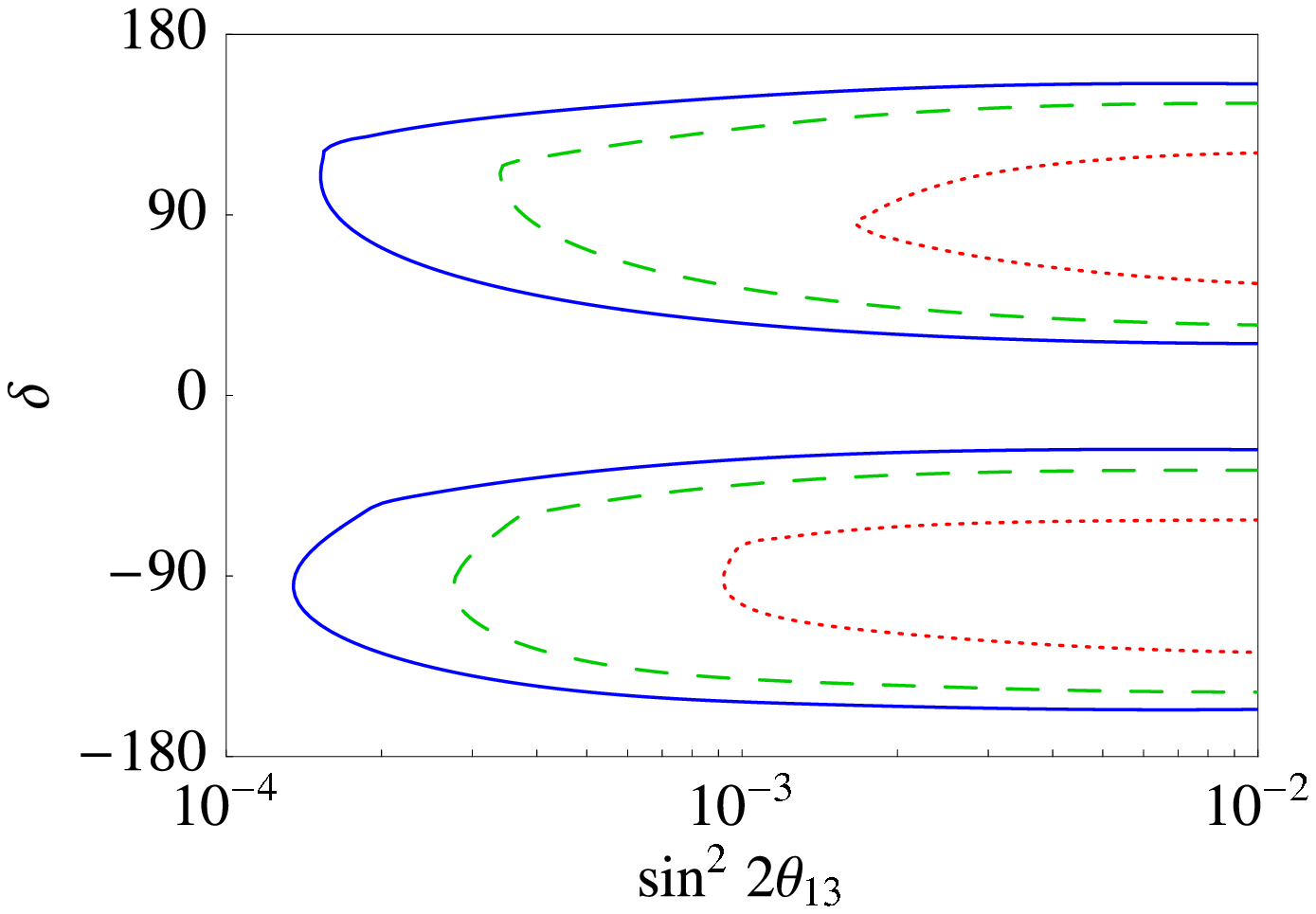} \\
\hspace{-0.55cm} \epsfxsize7.5cm\epsffile{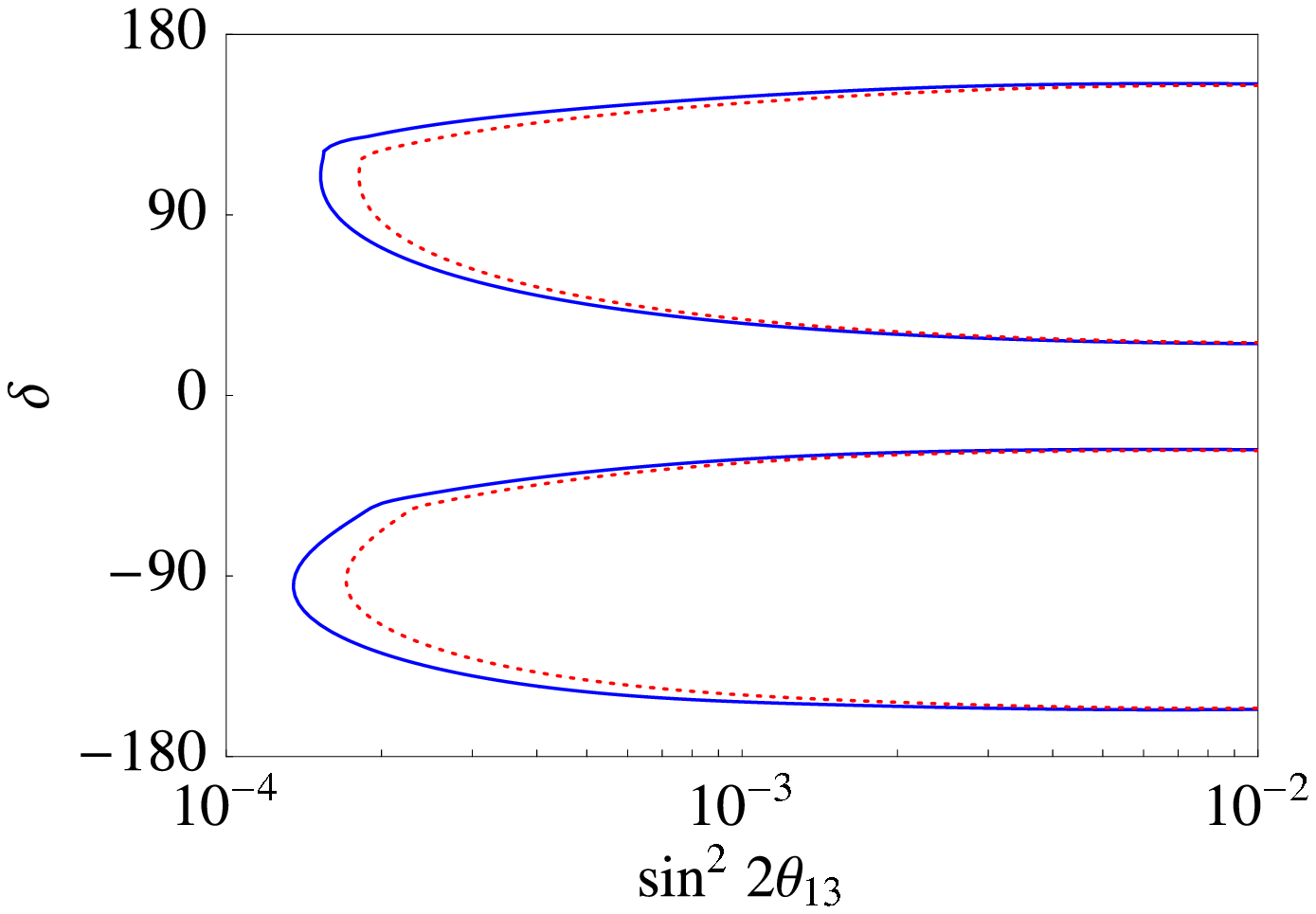} & 
                 \epsfxsize7.5cm\epsffile{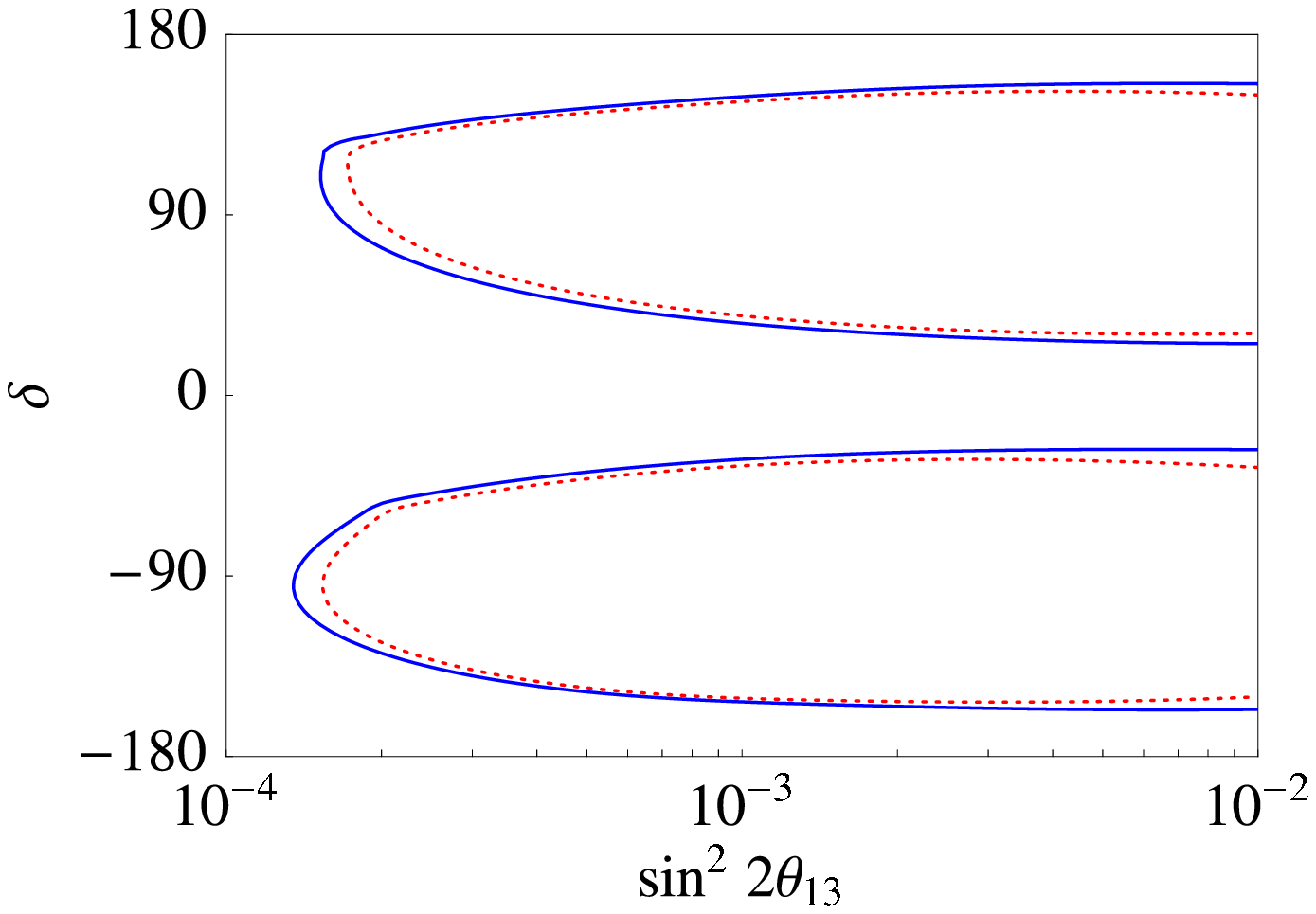}
\end{tabular}
\caption{\it 
3$\sigma$ CP discovery potential. 
Top left:  the impact of the flux for the $L = 2000$ Km baseline.
Dotted stands for $2 \times 10^{18}$, dashed for $5 \times 10^{18}$ and solid 
for $1  \times 10^{19}$ useful ion decays per year;
Top right:  the same, for the combination of the two baselines.
Bottom left: the impact of the beam-induced background for the combination of the two baselines
and a flux of $1 \times 10^{19}$ useful ion decays per year.
Dotted stands for a background of  $10^{-4}$ times the non-oscillated events, solid for
$10^{-5}$ times the non-oscillated events;
Bottom right: the impact of systematic errors for the combination of the two baselines and a flux of $1 \times 10^{19}$ useful ion decays per year. Solid stands for systematics of 2.5\% and 5\% on 
the signal and the background, respectively: dotted for systematics of 10\% and 20\% on 
the signal and the background, respectively. }
\label{fig:cpsens}
\end{center}
\end{figure}

In Fig.~\ref{fig:cpsens} we present our results for the CP discovery potential, defined in the following way:  the values of $\theta_{13}$ and $\delta$ in the plots represent the ``true'' values of these parameters. A ``true'' normal hierarchy is also assumed. 
For each of these input values, the $\chi^2$ for
$\delta=0^\circ$ and $\delta=180^\circ$ (marginalized in the rest of the parameters) were computed. If the value of the $\chi^2 > 9$, 
then the hypothesis that CP is conserved  can be rejected at 3$\sigma$ for those ``true'' values of $\theta_{13}$ and $\delta$.
Obviously this can never happen if the ``true'' value of $\delta$ is either $0^\circ$ or $180^\circ$, hence no sensitivity is found in stripes around those values of $\delta$.

In the top left panel we present the CP discovery potential for the three different fluxes at the 2000 Km baseline. Notice that for the ``nominal flux'', $2\times10^{18}$ useful ion decays per year (red dotted line), the low statistics at the detector and the presence of degeneracies at CP-conserving values
of $\delta$ spoil the discovery potential of the experiment. In this case there is no sensitivity to CP violation whatsoever. For the intermediate flux, $5\times10^{18}$ useful ion decays per year (green dashed line), some areas in which CP violation can be discovered appear. Sensitivity is again 
lost around $\sin^2 2\theta_{13}=4\times10^{-3}$ for negative $\delta$ and for 
$\sin^2 2\theta_{13}< 10^{-3}$ for positive $\delta$.  
Even for the ``ultimate flux'', $1 \times 10^{19}$ useful ion decays per year (blue solid line),
the CP discovery potential for negative values of $\delta$ around $\sin^22\theta_{13}=4 \times10^{-3}$ is lost. This is because sign degeneracies at $\delta = 180^\circ$ appear and do not allow to unambiguously determine CP violation, even if the true value of $\delta$ is CP-violating. This is the so called ``$\pi$-transit''  which also spoils the sensitivity of the $L \sim 3000$ Km detector 
of a Neutrino Factory for negative values of $\delta$ and $\sin^22\theta_{13}=3\times10^{-4}$  (see Fig.~8 of \cite{Huber:2002mx}).

However, as we will see in the next subsection, excellent sensitivities to the mass hierarchy can be achieved at the far detector observing the resonant  enhancement of the neutrino or antineutrino oscillation probability depending on whether the hierarchy is normal or inverted. 
The combination of the data taken at the two detectors can thus solve the sign degeneracy 
at $\pi$-transit and provide sensitivity to CP violation also for that region of the
parameter space. Moreover, at the $L = 7000$ Km, close to the ``magic baseline'', the effects of CP violation vanish providing a clean measurement  of $\theta_{13}$ that can greatly improve the CP discovery potential when combined with the data at 2000 Km. 
This combination is depicted in the top right panel of Fig.~\ref{fig:cpsens}, where now very good sensitivities to CP violation can be obtained for $\sin^22\theta_{13}>1.5 \times10^{-4}$.
Notice that we would get the same results for the combination of the two baselines in case
an inverted hierarchy were assumed.

In the bottom panels the impact of the beam-induced background (left) and 
of the systematic errors (right) on the CP discovery potential is studied, finding 
again that their effect is marginal.

\subsection{Sensitivity to the mass hierarchy}

\begin{figure}[t!]
\vspace{-0.5cm}
\begin{center}
\begin{tabular}{cc} 
\hspace{-0.55cm} \epsfxsize7.5cm\epsffile{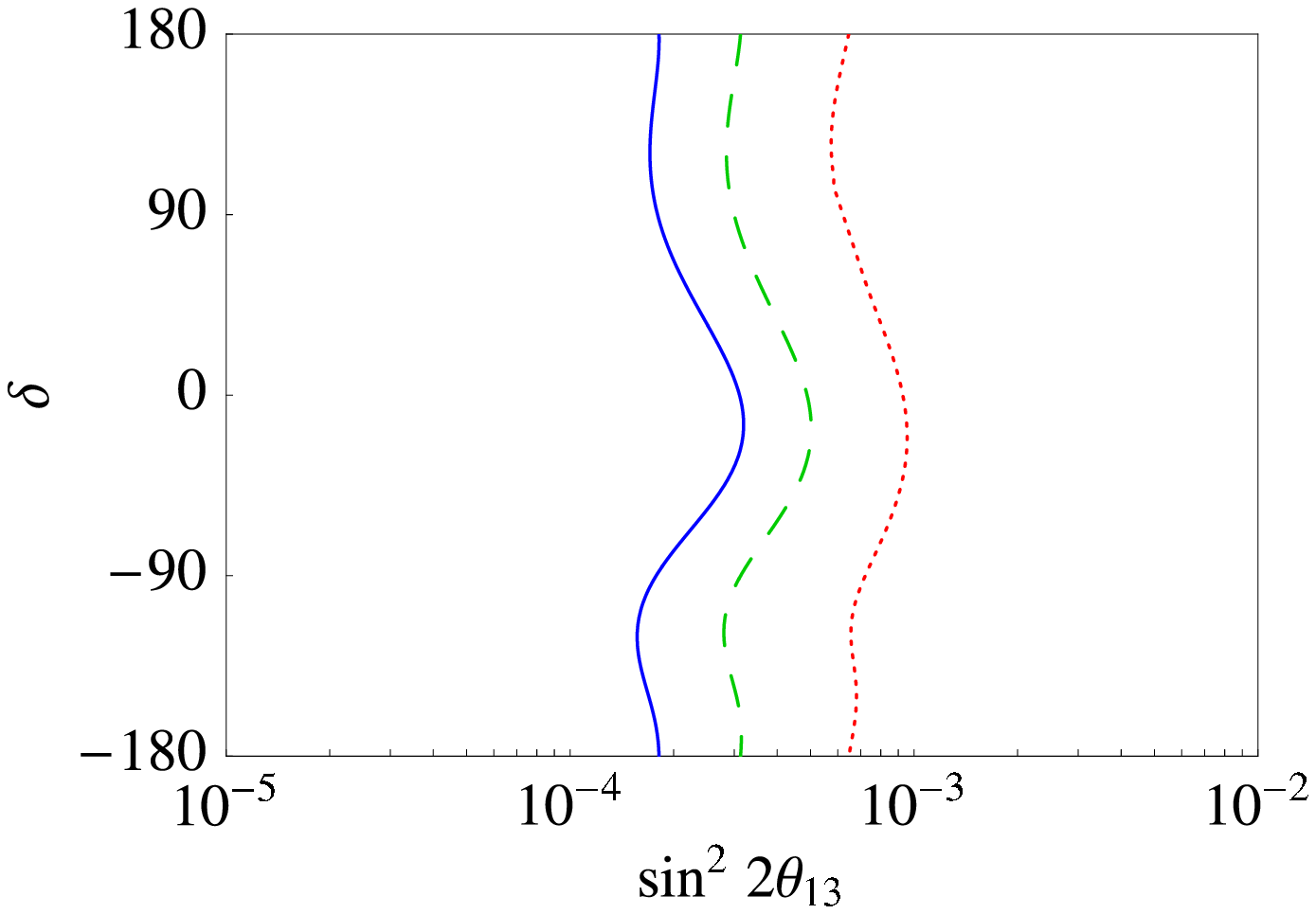} & 
                 \epsfxsize7.5cm\epsffile{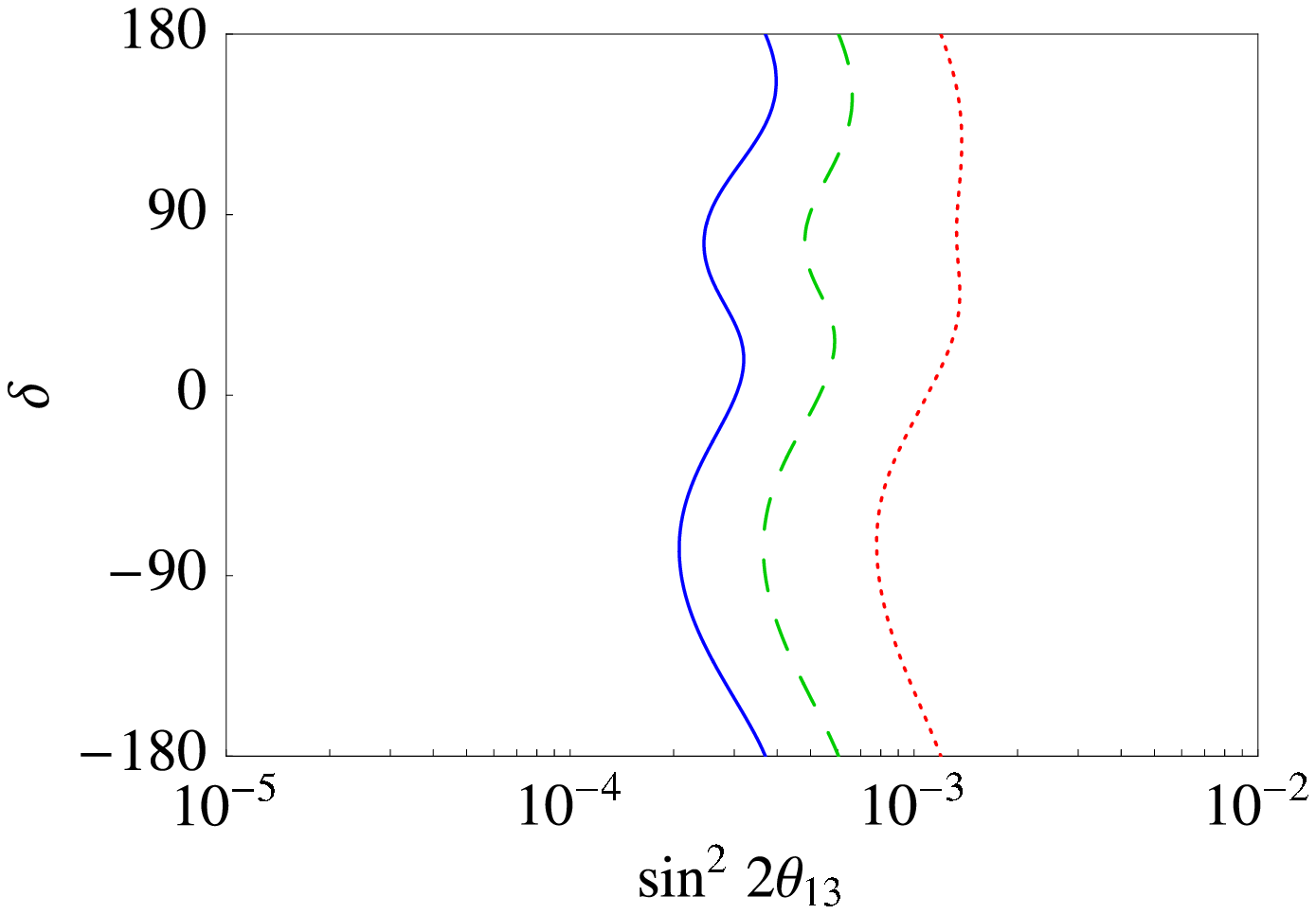} \\
\hspace{-0.55cm} \epsfxsize7.5cm\epsffile{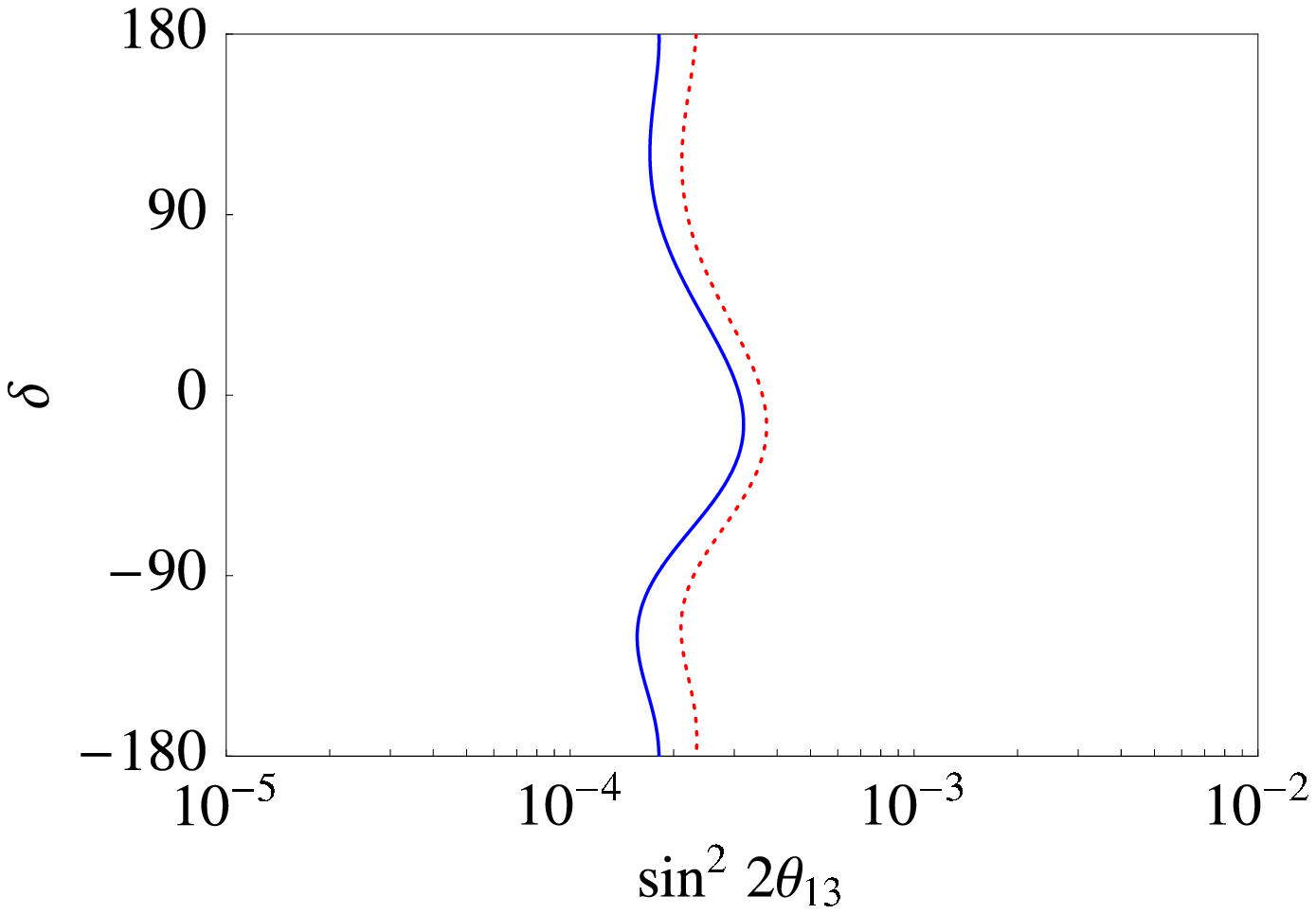} & 
                 \epsfxsize7.5cm\epsffile{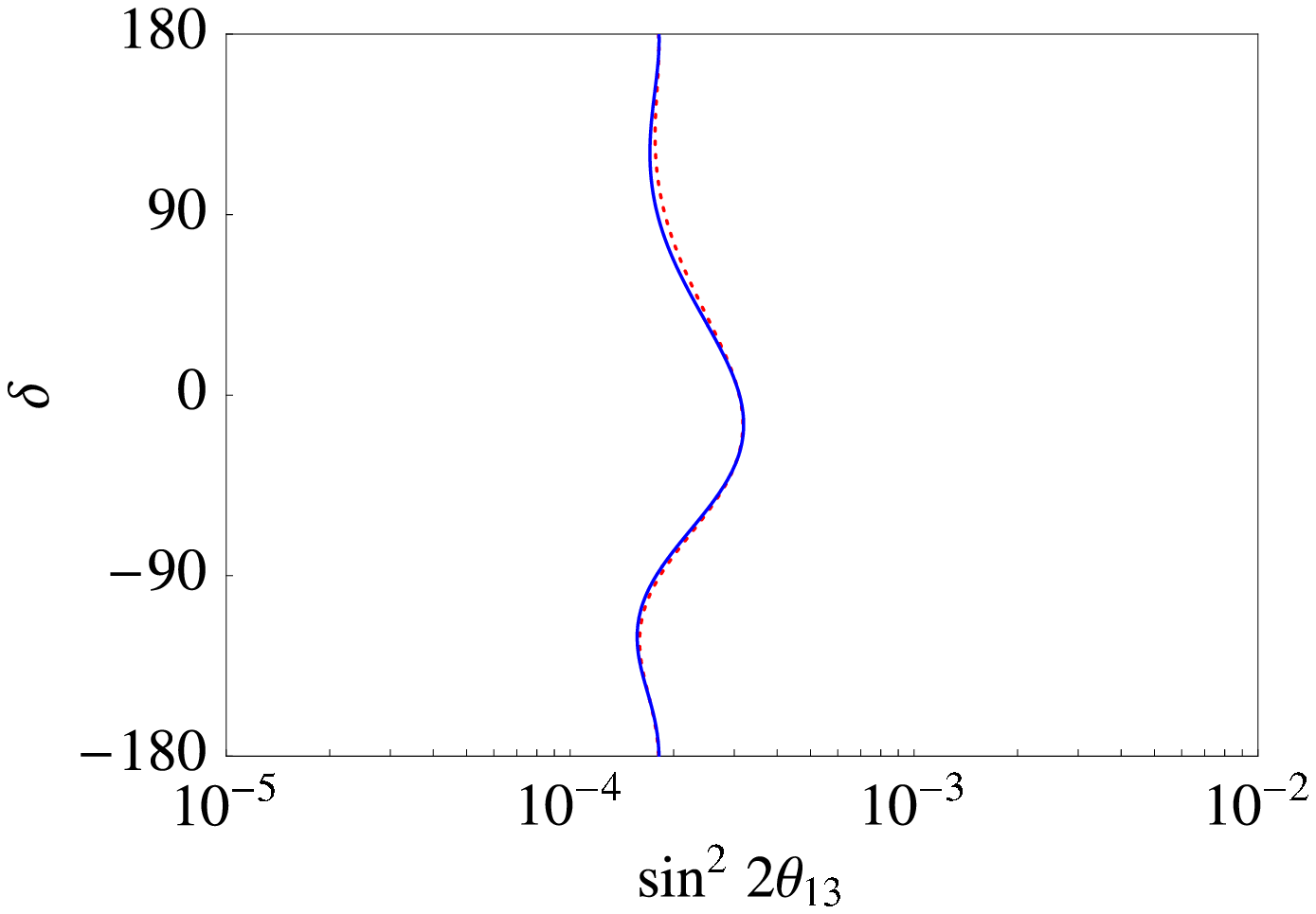}
\end{tabular}
\caption{\it 
3$\sigma$ sensitivity to the mass hierarchy. 
Top left: the impact of the flux for the combination of the two baselines and normal hierarchy.
Dotted stands for $2 \times 10^{18}$, dashed for $5 \times 10^{18}$ and solid 
for $1  \times 10^{19}$ useful ion decays per year;
Top right:  the same, for inverted hierarchy.
Bottom left: the impact of the beam-induced background for the combination of the two baselines
and a flux of $1 \times 10^{19}$ useful ion decays per year.
Dotted stands for a background of  $10^{-4}$ times the non-oscillated events, solid for
$10^{-5}$ times the non-oscillated events;
Bottom right: the impact of systematic errors for the combination of the two baselines and a flux of $1 \times 10^{19}$ useful ion decays per year. Solid stands for systematics of 2.5\% and 5\% on 
the signal and the background, respectively: dotted for systematics of 10\% and 20\% on 
the signal and the background, respectively. }
\label{fig:signsens}
\end{center}
\end{figure}

In Fig.~\ref{fig:signsens} we present our results for the sensitivity to the mass hierarchy, defined in the following way: 
the values of $\theta_{13}$ and $\delta$ in the plots represent the ``true'' values of these parameters. 
A given ``true'' hierarchy is also assumed. For each of these input values, the $\chi^2$
for the opposite mass hierarchy (marginalized in the rest of the parameters) was computed. If the value of the $\chi^2 > 9$, 
the wrong hierarchy can be rejected at 3$\sigma$ for those ``true'' values of $\theta_{13}$ and $\delta$.

In the top panels we present the sensitivity to the sign of the atmospheric mass difference
for the combination of the two baselines and three different fluxes, for normal (left)
and inverted (right) hierarchy. 
Notice that at 7000 Km either the neutrino or the antineutrino oscillation probability 
becomes resonant \cite{Agarwalla:2006vf,Agarwalla:2006gz}, depending on the mass hierarchy. 
As a consequence, the sensitivity to the sign of the atmospheric mass difference at this baseline is excellent: 
in Ref.~\cite{Agarwalla:2007ai}, indeed, sensitivity to $s_{atm}$ at 3$\sigma$ down to $\sin^2 2 \theta_{13} \geq 1 \times 10^{-3}$
(for $\gamma = 350$ and ``standard'' fluxes) is achieved. In our setup, due to the combination of the two baselines, a slightly better
sensitivity is at reach for ``nominal flux'', down to $\sin^22\theta_{13} = 8 \times 10^{-4} (1 \times 10^{-3})$ for normal (inverted) true hierarchy,  
whereas sensitivity down to $\sin^22\theta_{13}= 2 \times 10^{-4} (4 \times 10^{-4})$ is achievable for the ``ultimate flux''.
These sensitivities are enough to lift the sign degeneracy at the $\pi$-transit that causes
the loss of sensitivity to CP violation for negative values of $\delta$ (compare top left and top right panels in Fig.~\ref{fig:cpsens}). 

In the bottom panels, we again show the impact of the background (left) and of systematic errors (right), respectively.  
The effect of both is found to be very small also in this case. 

\subsection{Asymmetric detectors}

\begin{figure}[t!]
\vspace{-0.5cm}
\begin{center}
\begin{tabular}{ccc} 
\hspace{-0.55cm} \epsfxsize5cm\epsffile{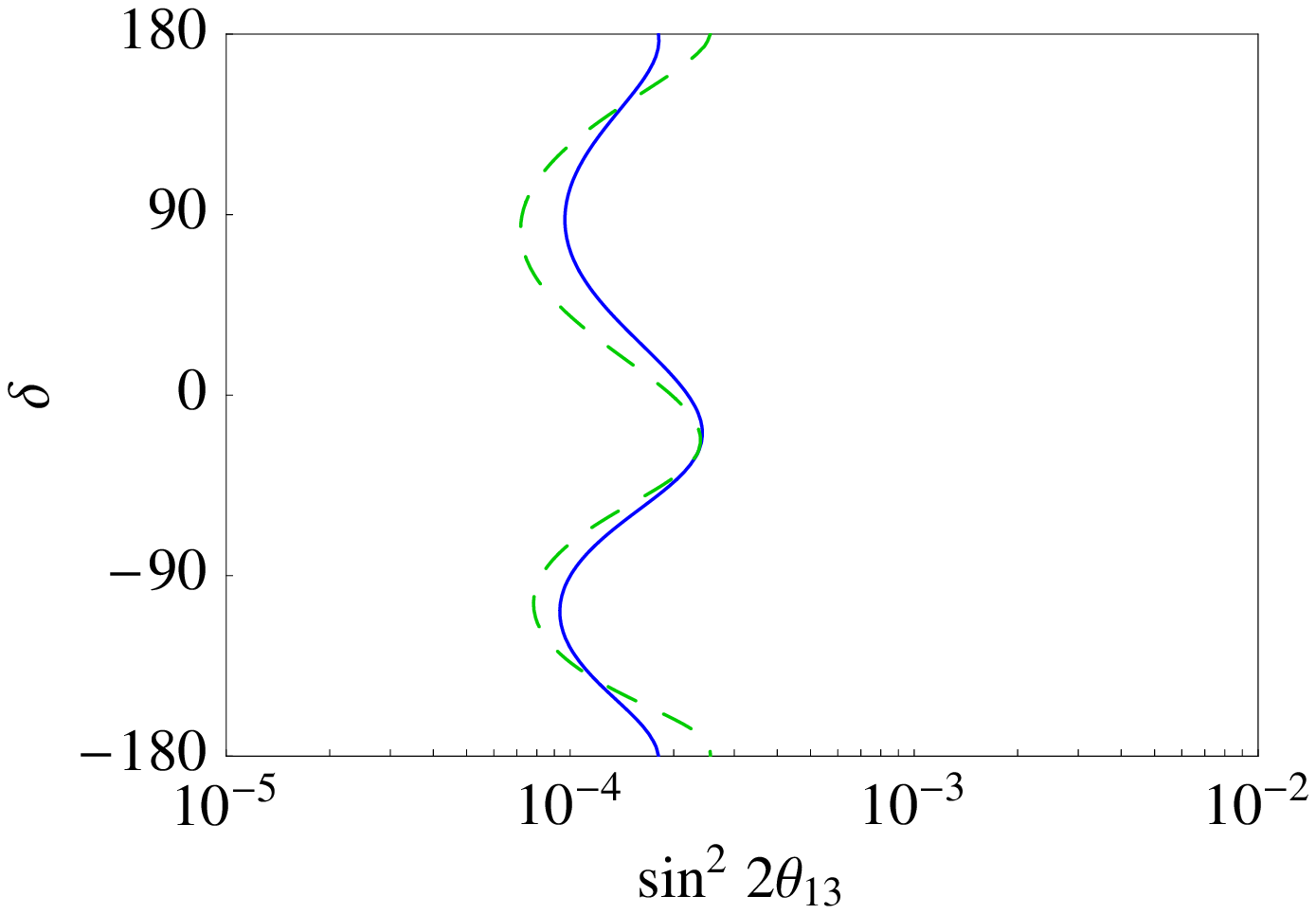} & 
                 \epsfxsize5cm\epsffile{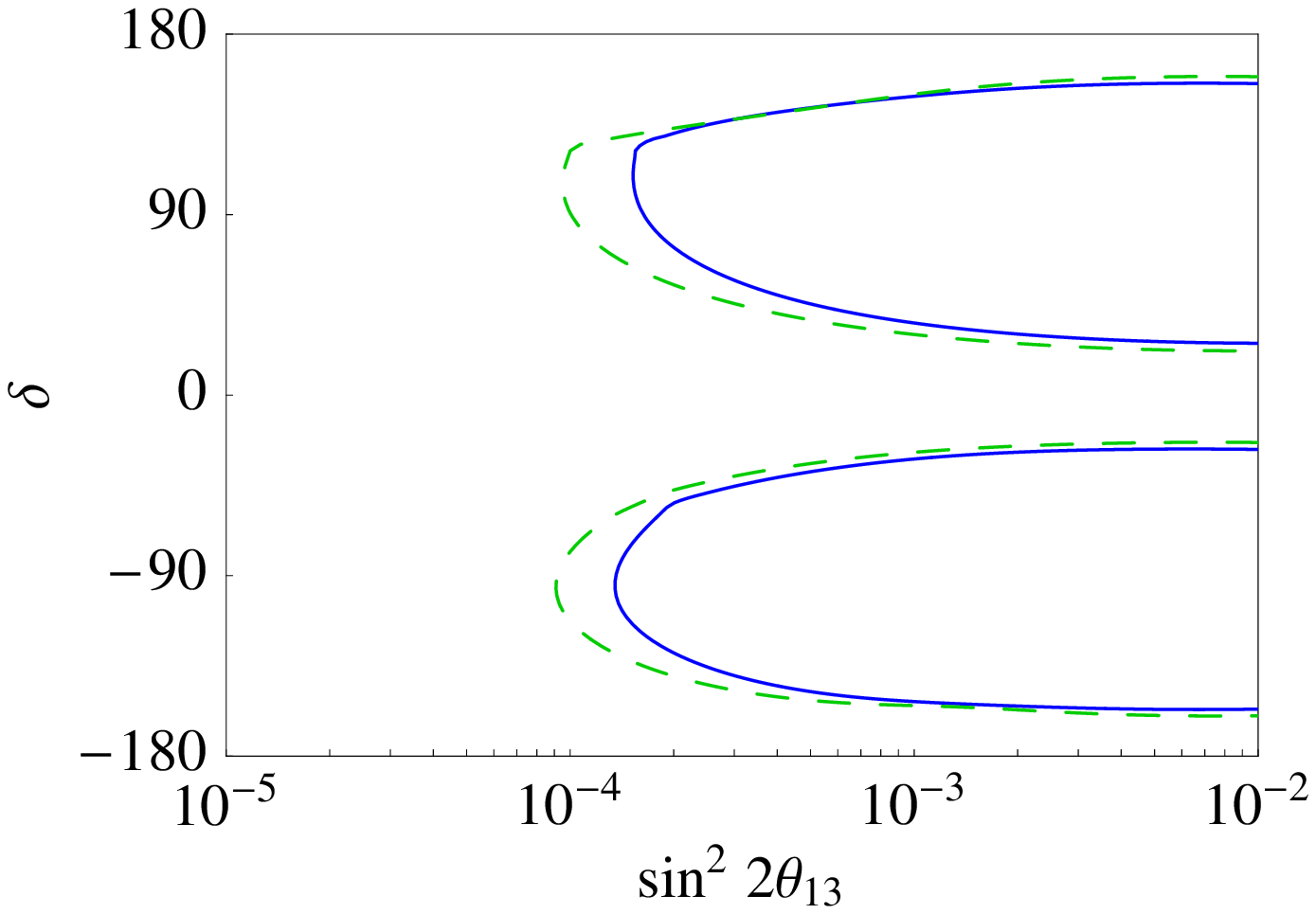} &
                 \epsfxsize5cm\epsffile{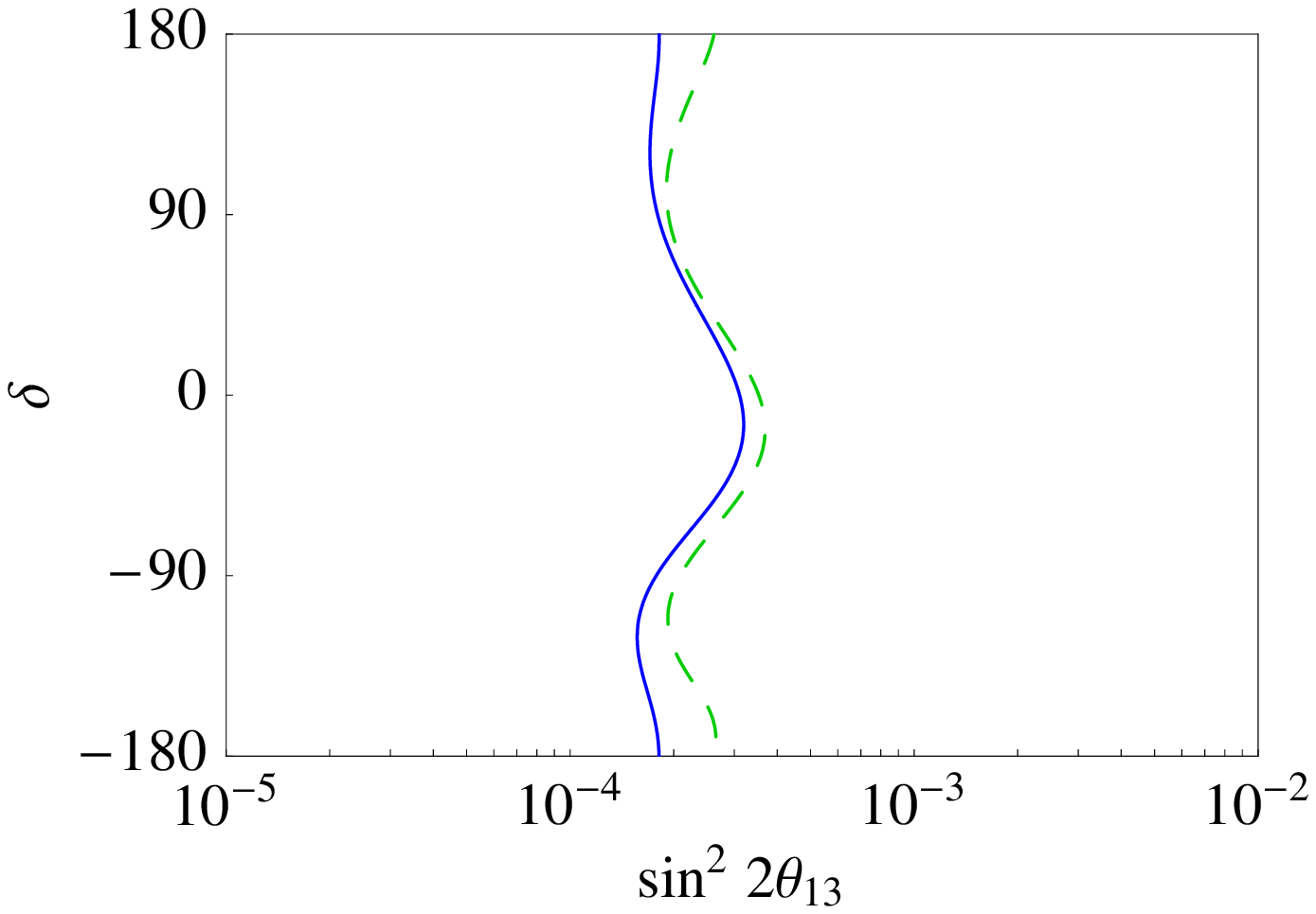}  
\end{tabular}
\caption{\it 
Comparison at 3$\sigma$ between the performance of two detectors with equal masses (50 Kton each) and an alternative 
setup considering an 80 Kton near detector and a 20 Kton far one. In all plots, solid lines correspond to the symmetric setup
and dashed lines to the asymmetric one. The left, middle and right panels show the $\theta_{13}$, $\delta$ and mass hierarchy discovery potentials, respectively. The results have been 
obtained considering a flux of $1 \times 10^{19}$ ion decays per year per straight section of the storage ring, a $10^{-5}$ fractional
background and a systematic error of 2.5\% on the signal and 5\% on the background.
}
\label{fig:asym}
\end{center}
\end{figure}

Up to this moment we have considered a symmetric setup in which two identical MIND-like 50 Kton detectors are located at $L = 2000$ Km and 
$L = 7000$ Km. The far detector is exploited, as explained in the Sect.~\ref{sec:pheno}, to solve the sign degeneracy in a CP-independent environment. 
To perform this task, however, it is not necessary to have such a large detector mass, due to the resonant effect in oscillation probabilities 
in matter for 6 GeV neutrinos at this baseline. 
At the price of losing some sensitivity to the hierarchy, thus, we can move some of the far detector mass to the near detector, increasing 
in this way the sensitivity to $\theta_{13}$ and $\delta$. 

This is shown in Fig.~\ref{fig:asym}, where we present our results for the sensitivity to $\theta_{13}, \delta$ and the mass hierarchy
for an asymmetric setup with a 20 Kton MIND-like far detector and an 80 Kton (otherwise identical) near one.
The left, middle and right panels show the $\theta_{13}$, $\delta$ and mass hierarchy discovery potentials, 
respectively.
The results have been obtained considering the ``ultimate'' flux, a $10^{-5}$ fractional background and a 
systematic error of 2.5\% on the signal and 5\% on the background. Solid lines stand for the symmetric 50 Kton case and dashed for the 80/20 Kton option.

As it was expected, we can see that the sensitivity to the hierarchy is slightly worse. In particular, we lose some sensitivity
for $\delta = 0,\pi$, going from $\sin^2 2 \theta_{13} \leq 2 \times 10^{-4}$ to $\sin^2 2 \theta_{13} \leq 3 \times 10^{-4}$. The sensitivity loss
for other values of $\delta$ is less significant. The same sensitivity loss for $\delta = 0,\pi$ is observed in the $\theta_{13}$ discovery potential. However, 
we can see that an increase in the $\theta_{13}$-sensitivity is achieved for $|\delta| = \pi/2$: we go from 
$\sin^2 2 \theta_{13} \leq 1 \times 10^{-4}$ to $\sin^2 2 \theta_{13} \leq 7 \times 10^{-5}$. This can be easily understood from the top left panel of Fig.~\ref{fig:t13sens}. The $\theta_{13}$ discovery potential at 2000 Km peaks for $|\delta| = \pi/2$ due to the increase in the neutrino (antineutrino) oscillation probability for $\delta = \pi/2$ ($\delta = -\pi/2$). On the other hand, at the magic baseline the $\delta$ dependence of the sensitivity is very mild and it is more strongly constraining $\theta_{13}$ near the CP-conserving values of $\delta$. The CP-violation discovery potential, depicted in the middle panel, improves for any value of $\delta$. In particular, 
for $|\delta| \sim \pi/2$ we go from $\sin^2 2 \theta_{13} \leq 1.5 \times 10^{-4}$ to $\sin^2 2 \theta_{13} \leq 9 \times 10^{-5}$.

Therefore, depending on the specific interest in a given physics observable, a symmetric setup or an asymmetric one should
be adopted. Adding mass to the near 
detector favors the sensitivity to CP-violation, whereas increasing the size 
of the far detector favors the mass hierarchy discovery potential.

\section{Conclusions}
\label{sec:concl}

We have tried to study a $\beta$-Beam setup capable of addressing the most relevant 
unmeasured neutrino oscillation parameters: $\theta_{13}$, the CP-violating phase $\delta$ 
and the neutrino mass hierarchy $s_{atm}$, solving the degeneracies among them. 
The setup proposed in this paper is a high-$\gamma$ ($\gamma=350$) $^8$Li/$^8$B $\beta$-Beam aimed at
two 50 Kton MIND-like magnetized iron detectors located at $L = 2000$ Km (on-peak) and $L = 7000$
Km (i.e. near the "magic" baseline) from the source. We have considered a constant 65\%
muon identification efficiency for the detectors, a $10^{-5}$ beam-induced fractional background
and 2.5\% and 5\% systematic errors on the signal and on the background, respectively. 
We have studied the sensitivity to $\theta_{13}$, $\delta$ and $s_{atm}$ of
this setup as a function of the flux (using $2 \times 10^{18}, 5 \times 10^{18}$ 
and $10 \times 10^{18}$ useful ion decays per year per polarity per baseline). 
We have also studied the dependence of the sensitivities on the beam-induced background and on the systematic errors. 

This setup is the natural development of a series of proposals that have been presented in the literature. 
On the technical side, we take advantage of: the ``ionization cooling'' technique to produce intense $^8$Li and $^8$B beams 
(the latter being out of reach with standard ISOL-type targets) \cite{Rubbia:2006pi}; the possible replacement of the SPS at CERN with a
new machine (the SPS+) \cite{PAF}, that would allow to boost ions up to $\gamma = 350$ \cite{Burguet-Castell:2003vv}; and, 
the study of a magnetized iron detector optimized for a multi-GeV neutrino beam \cite{cerveragolden07}.
On the phenomenological side, we have combined some of the features of the two most ambitious options 
identified in the ISS report, namely a double baseline ($L \sim 4000$ and $L \sim 7000$ Km, respectively), 
high flux ($10^{21}$ useful muons per year per polarity per baseline), high-energy ($E_\mu  \sim 30$ GeV) Neutrino Factory (NF-RS) \cite{Group:2007kx}, 
and the high-$\gamma$ $^6$He/$^{18}$Ne $\beta$-Beam aimed at a 1 Mton water \v Cerenkov located (on-peak) at $L = 650$ Km down the source 
\cite{Burguet-Castell:2003vv,Burguet-Castell:2005va}. 
The former has excellent sensitivity to $\theta_{13}$ and to the mass hierarchy (due to its long baselines and to the combination 
of the two detectors, that allows to measure $s_{atm}$ down to $\sin^2 2 \theta_{13} \geq 3 \times 10^{-5}$). The latter has 
excellent sensitivity to CP violation, outperforming the NF-RS for $\sin^2 2 \theta_{13} \geq 10^{-3}$. Since it is located on-peak 
and with very small matter effects that can mock true CP violation, it could detect a non-vanishing $\delta$ for more than $80 \%$ or even $90 \%$ 
of the parameter space if $\theta_{13}$ is not too small (see the right panel of Fig.~\ref{fig:optsummary}). 
The (too) short baseline, however, spoils its sensitivity to the mass hierarchy (see the left panel of Fig.~\ref{fig:optsummary}). 

The shorter baseline in our setup, $L = 2000$ Km, has thus been chosen so as to set the $\nu_e \to \nu_\mu$ oscillation at its first peak 
for the typical neutrino energies of the $\gamma = 350$ $^8$Li/$^8$B $\beta$-decays. Being on-peak helps in 
resolving the ($\theta_{13},\delta$) intrinsic degeneracy (a significant advantage with respect to the single baseline Neutrino
Factory setup). If only this detector were used, however, our results would still be strongly plagued by sign degeneracies because of 
its longer baseline with respect to the high-$\gamma$ He/Ne $\beta$-Beam. The combination of this baseline with the 7000 Km ``magic'' 
baseline, on the other hand, removes these degeneracies (alike the second baseline at the Neutrino Factory, \cite{Group:2007kx})
Moreover, as it was shown in Refs.~\cite{Agarwalla:2006vf,Agarwalla:2007ai}, the high-$\gamma$ Li/B
has huge resonant matter effects at the 7000 Km detector for energies around 5-6 GeV, that are missed by the (higher energy)
Neutrino Factory setups. 

The combination of the two baselines, thus, provides good sensitivity to the three observables. Notice that, if 
the ``medium'' or the ``ultimate flux'' can be achieved, this would be the only $\beta$-Beam--based setup capable of 
simultaneously probing CP violation and the neutrino mass hierarchy in the range $\sin^2 2\theta_{13} \in [3 \times 10^{-4}, 1 \times 10^{-2}]$. 
The main drawback is its low statistics compared to the He/Ne setup (with a shorter baseline and larger detector) or to the NF-RS
(whose flux is two orders of magnitude above the most optimistic ion flux considered here). 

Using the ``ultimate flux'' ($10 \times 10^{18}$), the sensitivity to $\theta_{13}$ is $\sin^2 2 \theta_{13} \geq 2\times 10^{-4}$, regardless
of the value of $\delta$. For specific values of $\delta$ close to maximal CP violation, $| \delta | \sim 90^\circ$, the sensitivity
reaches $\sin^2 2 \theta_{13} \geq 10^{-4}$, thus outperforming any Super-Beam or low-$\gamma$ $\beta$-Beam setup 
and being competitive with the $\gamma = 350$ He/Ne scenario (see Fig.~103 of Ref.~\cite{Group:2007kx}). 
For extremely small values of $\theta_{13}$ the NF-RS is, of course, unbeatable. 

It should be stressed that the comparison of our setup with ``ultimate flux'' with the high-$\gamma$ He/Ne $\beta$-Beam
is not completely fair, though, the ``ultimate flux'' being roughly five times larger than the ``standard flux'' used in the latter. 
However, we have shown that the $\theta_{13}$ sensitivity of our setup is limited by the statistical error, as it can 
be seen in Figs.~\ref{fig:t13sens}, \ref{fig:cpsens} and \ref{fig:signsens}. This is not the case for the high-$\gamma$ He/Ne $\beta$-Beam, 
whose sensitivity is dominated by systematics and backgrounds (notice that the statistics at this setup is, with ``standard flux'', 
approximately ten times larger than for the high-$\gamma$ Li/B $\beta$-Beam with ``ultimate flux'', see Ref.~\cite{Burguet-Castell:2003vv}). 
Notice, eventually, that the suppression factor $S_f$ needed to reduce  to acceptable levels the atmospheric muon backgrounds
is much smaller for the high-$\gamma$ Li/B $\beta$-Beam with respect to low- and high-$\gamma$ He/Ne beams, due to the higher
average neutrino energy (see Sect.~\ref{sec:det}). 
For fixed bunch time-length $\Delta t_{bunch} = 10$ ns, therefore, a higher number of bunches can be injected into the storage ring, 
thus making an increase in the flux easier. 

The CP-violating phase $\delta$ can be measured in approximately $70\%$ of the $\delta$ parameter space for $\sin^2 2\theta_{13} \geq 10^{-3}$. 
Some sensitivity to $\delta$ is achieved for $|\delta| \sim 90^\circ$ down to $\sin^2 2\theta_{13} \geq 10^{-4}$.
Comparing this with Fig.~105 of Ref.~\cite{Group:2007kx}, we again find that this setup outperforms all Super-Beams and low-$\gamma$
$\beta$-Beam scenarios. It is, however, outperformed by the high-$\gamma$ He/Ne $\beta$-Beam (that covers around a 85\% 
of the $\delta$ parameter space down to $\sin^2 2\theta_{13} \geq 2 \times 10^{-3}$, with some sensitivity down to $\sin^2 2\theta_{13} \geq 
5 \times 10^{-5}$ for $|\delta| \sim 90^\circ$) and by the NF-RS (that covers around a 80\% 
of the $\delta$ parameter space down to $\sin^2 2\theta_{13} \geq 2 \times 10^{-4}$, with some sensitivity down to $\sin^2 2\theta_{13} \geq 
2 \times 10^{-5}$ for $|\delta| \sim 90^\circ$). 

As for the sensitivity to the mass hierarchy, we find that the true hierarchy could be identified if 
$\sin^2 2\theta_{13}\geq 3 \times 10^{-4}$ for any value of $\delta$, with some sensitivity down to     
$\sin^2 2\theta_{13}\geq 10^{-4}$ for $|\delta | \sim 90^\circ$. Compared with Fig.~104 of Ref.~\cite{Group:2007kx}, only the NF-RS 
(with sensitivity down to $\sin^2 2\theta_{13}\geq 2 \times 10^{-5}$) can beat the high-$\gamma$ Li/B $\beta$-Beam, 
the rest of the considered facilities being between one and two orders of magnitude worse.

Depending on the specific interest in a given physics observable the mass ratio of the near and far detectors can be varied. Increasing the mass of the near 
detector improves the sensitivity to CP-violation, whereas increasing the size 
of the far detector improves the mass hierarchy discovery potential.

In summary, we think that the combination of the ``on peak'' and ``magic'' baselines at a high-$\gamma$ Li/B $\beta$-Beam
is a very powerful tool to solve degeneracies and find good sensitivities to the most relevant unknown parameters
of the leptonic flavour sector. This setup is, however, limited by the statistical error and would strongly benefit of any improvement 
on the neutrino flux, detector mass or efficiency.

\section*{Acknowledgments}

We thank C.~Biggio, A.~Cervera, S.~Choubey, J.J.~G\'omez-Cadenas, P.~Hern\'andez, M.~Lindroos, A.~Lombardi, P.~Migliozzi, S.~Rigolin and O.~Tengblad 
for useful discussions. The authors acknowledge the financial support of Ministerio de Educaci\'on y Ciencia (MEC) through project FPA2006-05423, 
of the Comunidad Aut\'onoma de Madrid through project HEPHACOS P-ESP-00346 and of the European Union through the networking activity BENE, 
RII3-CT-2003-506395. 
E.F.M. acknowledges financial support from the Universidad Aut\'onoma de Madrid and from the Max Planck Institut f\"ur Physik. J.L.P. 
acknowledges financial support from the MEC through the FPU grant AP2005-1185.

\begin{appendix}

\section{The acceleration and storage chain}
\label{sec:app}

\begin{figure}
\centering
\epsfig{file=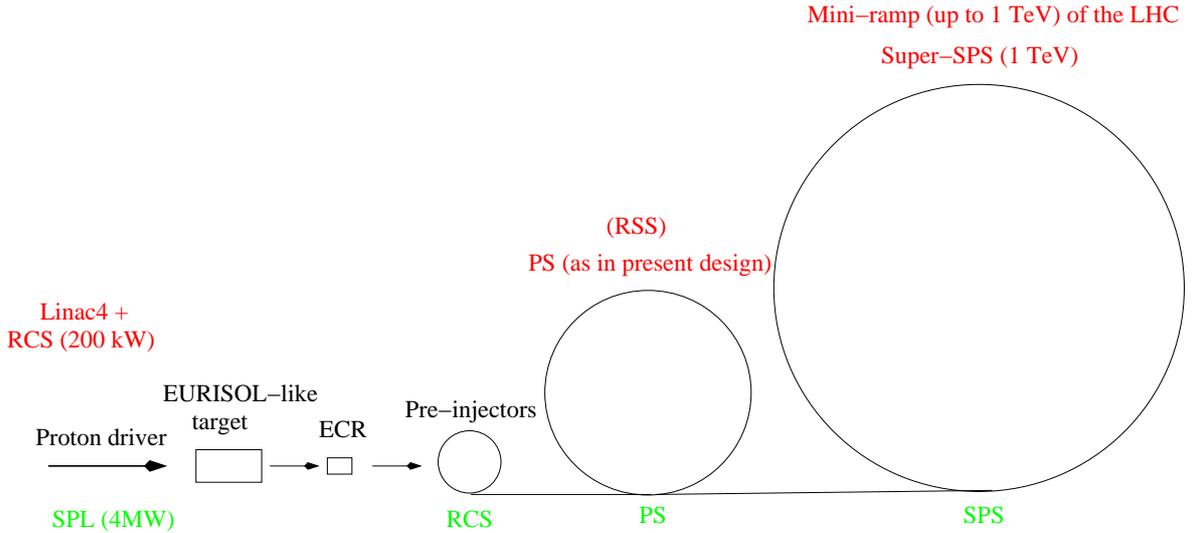,width=\textwidth}
\caption{The main components of the $\beta$-Beam production and acceleration complex.
In the lower part, the machines considered in the baseline option are indicated
(where RCS stands for Rapid Cycling Synchrotron). The alternatives that profit of the
LHC maintenance and upgrade programme are mentioned in the upper part.
The Rapid Superconducting Synchrotron~\cite{garoby_hif04} (RSS) is a possible upgrade of the PS. 
Eventually, the Super-SPS is presently known as ``SPS+''.}
\label{fig:complex}
\end{figure}

We show in Fig.~\ref{fig:complex} a schematic view of the infrastructure needed to produce and 
accelerate the ion beam. Notice that this scheme was designed for ``standard'' $^6$He/$^{18}$Ne beams.
The main difference with respect to our scheme, in which $^8$Li/$^8$B ions are produced, is the target. 
Instead of an EURISOL target station, we need a device in which produce a sustained $^8$Li/$^8$B
flux using the ``ionization cooling'' technique. For details on the ion production, see Refs.~\cite{Rubbia:2006pi,betabeams_moriond}.
In the rest of this Appendix we will make some comment on differences and similarities of the He/Ne setups (for $\gamma = 100$ or 350) 
with the one used in this paper. Numbers quoted here are taken from the EURISOL collaboration webpage~\cite{Lindroos,Wildner:2007dy}. 
They have been computed for $^6$He and $^{18}$Ne ions boosted at $\gamma = 100$, 
trying to achieve the goal luminosity of $2.9 \times 10^{18}$ $^6$He 
and $1.1 \times 10^{18}$ $^{18}$Ne ion decays per year, respectively.

\subsection{Proton Driver}

In the baseline design, the proton driver is the proposed Super Proton Linac (SPL), a multi-MW ($\sim 4$~MW, $E_p=$2.2~GeV~\cite{SPL} 
or 3.5~GeV~\cite{Garoby:2005se,Campagne:2004wt}) machine aimed at substituting the present Linac2 and PS Booster (PSB). 
Contrary to naive expectation, however, a multi-MW booster is not needed for the construction of a $\beta$-Beam 
or an EURISOL facility\footnote{The SPL (or a similar proton driver) is mandatory, instead, for a low-energy neutrino SuperBeam~\cite{Mezzetto:2003mm} 
or for a Neutrino Factory.}.
Any of the possibilities currently under discussion at CERN for the upgrade of the PSB, based either on Rapid Cycling Synchrotron's or on Linac's, 
represents a viable solution for the production stage of a $\beta$-Beam complex. 
In the framework of the LHC maintenance and upgrade programme, the PAF committee \cite{PAF} suggested the substitution of the Linac2
with a new Linac (Linac4) that will inject protons into the PSB at 160 MeV. 
This would allow production of $\sim 2{\times} 10^{13}$ $^6$He/s for 200~kW on target, consistent with the current SPL-based design.  

\subsection{Ion production}
\label{sec:ionprod}

According to the latest numbers by the EURISOL collaboration~\cite{Lindroos}, $5 \times 10^{13}$ $^6$He atoms/s and $2 \times 10^{12}$ $^{18}$Ne atoms/s
can be produced using standard BeO and MgO ISOL targets, respectively. No relevant changes in these numbers are expected using a different design 
for the proton driver. The $^{18}$Ne production rate is still low to obtain the goal luminosity $1.1 \times 10^{18}$ $\nu_e$/year.

Once isotopes are produced, they are collected and ionized using an ECR ion source. 
Ionization efficiency at this stage is $^{18}$Ne is 29\%, whereas for the $^6$He flux is ~93\% \cite{Lindroos}. 

\subsection{Acceleration stage}

\begin{table}[htb]
\begin{center}
\begin{tabular}{|l|c|c|c|c|}
\hline
\noalign{\smallskip}
    & \textbf{$^6$He} & \textbf{$^{18}$Ne} & \textbf{$^8$Li} & \textbf{$^8$B} \\
\hline\hline
\noalign{\smallskip}
\textbf{ LINAC} &    &    &    &  \\[1mm]
(Injected ions/s)/$10^{12}$ & 17.1 \cite{Lindroos} & 5.25 \cite{Lindroos} & 17.1 & 17.1 \\[1mm]
$\gamma_{final}$ & 1.11 \cite{Lindroos} & 1.11 \cite{Lindroos} & 1.10 & 1.10\\[1mm]
\hline\hline
\noalign{\smallskip}
\textbf{RCS} ($L = 251$ m) &    &    &    &  \\[1mm]
(Injected ions/s)/$10^{12}$ & 8.53 \cite{Lindroos} & 2.62 \cite{Lindroos} & 8.53 & 8.53 \\[1mm]
$\gamma_{final}$ & 1.84 \cite{Lindroos} & 2.77 \cite{Lindroos} & 1.9 & 2.9 \\[1mm]
P (KW) &  $\lesssim$ 0.1 & $\lesssim$ 0.1 & $\lesssim$ 0.1 & $\lesssim$ 0.1 \\[1mm]
$t_{acc} (s)$ & 0.0475 \cite{Lindroos} & 0.0475 \cite{Lindroos} & 0.0475 & 0.0475 \\[1mm]
\hline\hline
\noalign{\smallskip}
\textbf{PS} ($L = 628$ m) &    &    &    &  \\[1mm]
(Injected ions/s)/$10^{12}$ & 1.84 \cite{Lindroos} & 1.25 \cite{Lindroos} & 1.84 & 1.84 \\[1mm]
$\gamma_{final}$ & 9.33 \cite{Lindroos} & 15.53 \cite{Lindroos} & 10.3 & 17.4 \\[1mm]
P (KW)       &  0.8 & 1.0 & 1.1 & 1.3 \\[1mm]
P/L (W/m)    &  1.3 & 1.6 & 1.8 & 2.1 \\[1mm]
$t_{acc} (s)$ & 0.8 \cite{Lindroos} & 0.8 \cite{Lindroos} & 0.8 & 0.8 \\[1mm]
\hline\hline
\noalign{\smallskip}
\textbf{SPS} ($L = 6912$ m)&    &    &    &  \\[1mm]
(Injected ions/s)/$10^{12}$ & 1.59 \cite{Lindroos} & 1.20 \cite{Lindroos} & 1.61 & 1.68 \\[1mm]
$\gamma_{final}$ & 100 & 100 & 100 & 100 \\[1mm]
P (KW)    & 3.0 & 1.8 & 3.4 & 2.2 \\[1mm]
P/L (W/m) & 0.4 & 0.3 & 0.5 & 0.3 \\[1mm]
$t_{acc} (s)$ & 2.54 \cite{Lindroos} & 1.42 \cite{Lindroos} & 2.2 & 1.2 \\[1mm]
\hline
\end{tabular}
\caption{Relevant beam parameters at the different acceleration stages of the standard $\beta$-Beam setup for $^6$He/$^{18}$Ne and $^8$Li/$^8$B.}
\label{tab:accstage}
\end{center}
\end{table}

In Tab.~\ref{tab:accstage} we give the relevant parameters, extracted from Ref.~\cite{Lindroos}, for the acceleration stage of He/Ne ions
up to $\gamma = 100$ in the standard (reference) setup. At the same time, we compute the values of the same parameters for Li/B ions
injected and accelerated using the same setup up to $\gamma = 100$. We assume for Li/B the same number of ions injected per second
in each stage as for He ions. Notice that no asymmetry is expected for Li/B, since both ions have similar $Z$, $A$ and $T_{1/2}$ (see Tab.~\ref{tab:ions}). 
This is not the case for $^6$He and $^{18}$Ne, whose production and ionization stages differ significantly. 

Unfortunately, we cannot perform a similar computation of these parameters
for the setup actually used in this paper (i.e. accelerating Li/B ions up to $\gamma = 350$) in the absence of a detailed technical 
specification of the acceleration chain. 

The standard acceleration stage consists of four steps: 
\begin{enumerate}
\item 
Ions are accelerated to $\gamma = 1.1$ introducing them into a LinAc (where $^{18}$Ne ions get fully ionized). 
\item 
They are then injected into the Rapid Cycling Synchrotron (RCS), where they reach 
$\gamma_{^6He} = 1.8$ and $\gamma_{^{18}Ne} = 2.8$ ($\gamma_{^8Li} = 1.9$ and $\gamma_{^8B} = 2.9$).
The transfer efficiency from the LinAc to the RCS is 50\%, only.
The different $\gamma$ reached at the end of this phase depends on the different $A/Z$, 
at fixed magnetic field, radius and acceleration time ($t_{acc} \sim 0.05$ s). 
No significant power losses (see, e.g., Ref.~\cite{Wildner:2007dy}) are expected at this stage ($\lesssim 0.1$ KW), for any of the considered ions.
No benefit from the LHC maintenance and upgrade phase is expected.

\item 
After the RCS, ions are transported and injected into the PS. 
Once in the PS, they are accelerated up to $\gamma_{^6He} = 9.3$ and $\gamma_{^{18}Ne} = 15.5$ ($\gamma_{^8Li} = 10.3$ and $\gamma_{^8B} = 17.4$),
in $t_{acc} \sim 0.8$ s.
Significant power losses are expected at this stage, for any of the considered ions\footnote{
We have computed the number of lost particles and the dissipated power at each acceleration stage following Ref.~\cite{Wildner:2007dy}, 
getting good agreement with Ref.~\cite{Lindroos} for the RCS and the SPS and for the particle loss at the PS. 
However, our result for the power loss for $^6$He and $^{18}$Ne in the PS differ from that reported in Ref.~\cite{Lindroos}. 
We do not understand the source of the disagreement. 
Notice that our conclusion is not affected by this discrepancy, though: 
the total dissipated power at the PS exceeds presently allowed values.}: 
$P_{^6He} \simeq 0.8$ KW, $P_{^{18}Ne} \simeq 1.0$ KW, $P_{^8Li} \simeq 1.1$ KW, $P_{^8B} \simeq 1.3$ KW.
This corresponds to a power loss per meter of $P_{^6He}/L_{PS} = 1.3$ W/m, $P_{^{18}Ne}/L_{PS} = 1.6$ W/m, 
$P_{^8Li}/L_{PS} = 1.8$ W/m, $P_{^8B}/L_{PS} = 2.1$ W/m. These values exceed the permitted upper limit, 1 W/m. 
This is a well known problem, see e.g. Ref.~\cite{Autin:2002ms} and \cite{Lindroos}, that must be solved if we are to use the PS at CERN
as a second stage ion beam accelerator. 

This acceleration stage could greatly benefit of the LHC maintenance and upgrade programme. 
The PS is the oldest machine in the CERN complex, and it has been proposed to replace it with 
a new 50 GeV synchrotron (called ``PS2'') \cite{PAF}. Using the PS2, substantial improvements are expected. 

\item 
After coming out of the PS, ions are transported and injected into the SPS with no significant expected losses.
In this last acceleration stage, they are boosted to $\gamma = 100$ in $t_{^6He} = 2.5$ s and $t_{^{18}Ne} = 1.4$ s 
($t_{^8Li} = 2.2$ s and $t_{^8B} = 1.2$ s).
Power losses at this stage are: $P_{^6He} = 3.0$ KW, $P_{^{18}Ne} = 1.8$ KW, $P_{^8Li} = 3.4$ KW and $P_{^8B} = 2.2$ KW.
In this case, due to the longer SPS circumference ($L = 6912$ m), the power loss per meter is:
$P_{^6He}/L_{SPS} = 0.4$ W/m, $P_{^{18}Ne}/L_{SPS} = 0.3$ W/m, $P_{^8Li}/L_{SPS} = 0.5$ W/m and $P_{^8B}/L_{SPS} = 0.3$ W/m,
well within acceptable limits for all considered ions.

The SPS+ would have an enormous impact on the design of a $\beta$-Beam at CERN. 
A detailed simulation of the acceleration and losses at this facility for any of the considered ions is lacking.
Notice, however, that the proposed design fulfills simultaneously the two most relevant requirements for a high energy $\beta$-Beam booster: 
it provides a fast ramp ($dB/dt=1.2 \div 1.5$~T/s~\cite{fabbricatore}) to minimize the number of decays during the acceleration phase 
and it can increase the $\gamma$ boosting factor up to $\gamma \leq 600$. 

\end{enumerate}

The outcome of our analysis is that accelerating He/Ne or Li/B ions using the same acceleration chain up to the same final $\gamma$
will give similar results for particle losses and dissipated power. 

\subsection{Storage ring}
\label{sec:storage}

The increase of the ions energy in the last element of the booster chain represents 
a challenge~\cite{terranova_nufact04}. Ions of high rigidity must be collected in a dedicated ring of reasonable size. 
The ring should have long straight sections ($L_{straight}=2500$~m) oriented alongside the direction of 
the far detector, with as small as possible curved sections (``racetrack'' geometry). 
In the baseline design with 5 T magnets \cite{Zucchelli:sa}, the radius of the curved sections is $R\sim 300$~m.

Decays that provide useful $\nu_e/\bar \nu_e$ are those occurring in the straight section for which ions travel toward
the detector. The useful fraction of ion decays (also called ``livetime") is defined as: 
\begin{equation}
\nonumber
l_{racetrack} = \frac{L_{straight}}{2 L_{straight} + 2 \pi R}
\end{equation}
For the baseline design, since $L_{ring} = 2 L_{straight} + 2 \pi R = 6911$ m (i.e., as long as the SPS), 
the livetime is $l^{100}_{He/Ne} = 0.36$ \cite{Zucchelli:sa}. 

A decay ring with straight sections of the same length, equipped with LHC 8.3 T dipolar magnets, 
 would stack ions boosted at $\gamma = 350$ at the nominal 
SPS+ rigidity with a significantly larger radius ($\sim 600$~m). The corresponding livetime is, thus, $l^{350}_{He/Ne} = 0.28$. 
This means that larger ion fluxes are needed to achieve the goal $\nu_e/\bar \nu_e$ luminosities\footnote{
Current R\&D related with the LHC upgrades and aimed at the development of high field magnets
(11$\div$15~T)~\cite{bruning,Devred:2005vd} could be used to increase the livetime.}.  

Our setup differs from both the baseline setup and the high-$\gamma$ 
$^6$He/$^{18}$Ne setup. First of all, for a fixed magnetic field a somewhat shorter radius is needed to bend the  $^8$Li/$^8$B beams 
with respect to $^6$He/$^{18}$Ne beams, due to the different $A/Z$ ($R_{Li/B} = 533$ m for $\gamma = 350$).
As a consequence, a livetime $l^{350}_{Li/B} = 0.30$ is obtained for a ring with a racetrack geometry.
However, since we propose to illuminate two detectors located at different baselines
at the same time,  two racetrack geometry storage rings should be built. 

An alternative option is to consider a different storage ring geometry. For a triangular
storage ring (as the one proposed also for some Neutrino Factory setups), we have:
\begin{equation}
\nonumber
l_{triangle} = \frac{2 L_{straight} }{(2 \pi R + 3 L_{straight}) }
\end{equation}
For three straight sections $L_{straight} = 2500$ m and three arcs with $R = 533$ m curvature radius, 
we have a livetime $l^{350}_{Li/B} = 0.46$. Notice that with this geometry the storage ring is 
$L_{ring} \sim 10$ Km, not much larger than in the racetrack case. 
Eventually we remind that, in this case, 23\% of the ions circulating in the ring produce the neutrino beam aimed at each of the two detectors. 
This must be compared with the single detector setup with racetrack storage ring, where 30\% of the ions produce a useful neutrino beam. 

Activation of the ring in the baseline setup ($^6$He/$^{18}$Ne at $\gamma = 100$) is under study. 
Results presented at the last NuFact Conference seem to indicate that energy deposit 
at the end of the straight sections is under control \cite{FabichNuFact07}.
Power losses and activation for $^6$He/$^{18}$Ne at higher $\gamma$ have not been computed in detail, however.

In the case of $^8$Li and $^8$B ion beams, no detailed study has been performed for any $\gamma$.
The $\beta^-$-decay channel of $^8$Li is $^8$Li $\to$ $^8$Be $\to$ 2$\alpha$. 
$^8$B also decays into $^8$Be and finally to two $\alpha$'s (it is the mirror nucleus of $^8$Li).
The two $\alpha$'s have the same $A/Z$ as the $^6$Li, and therefore the energy deposit
should be located in the same part of the magnets at the end of the ring straight sections.
 It must be reminded that the $^8$B $\beta$-decay spectrum is affected by several 
systematics errors that must be tamed before using it for a precision experiment (see Ref.~\cite{Winter:2003ac}).

\end{appendix}


\begin{thebibliography}{999}

\bibitem{Cleveland:1998nv}
  B.~T.~Cleveland {\it et al.},
  Astrophys.\ J.\  {\bf 496} (1998) 505.

\bibitem{Abdurashitov:1999zd}
  J.~N.~Abdurashitov {\it et al.}  [SAGE Collaboration],
  Phys.\ Rev.\  C {\bf 60} (1999) 055801
  [arXiv:astro-ph/9907113].

\bibitem{Hampel:1998xg}
  W.~Hampel {\it et al.}  [GALLEX Collaboration],
  Phys.\ Lett.\  B {\bf 447} (1999) 127.

\bibitem{Fukuda:2001nj}
  S.~Fukuda {\it et al.}  [Super-Kamiokande Collaboration],
  Phys.\ Rev.\ Lett.\  {\bf 86} (2001) 5651
  [arXiv:hep-ex/0103032].

\bibitem{Ahmad:2001an}
  Q.~R.~Ahmad {\it et al.}  [SNO Collaboration],
  Phys.\ Rev.\ Lett.\  {\bf 87} (2001) 071301
  [arXiv:nucl-ex/0106015].

\bibitem{Ahmed:2003kj}
  S.~N.~Ahmed {\it et al.}  [SNO Collaboration],
  Phys.\ Rev.\ Lett.\  {\bf 92} (2004) 181301
  [arXiv:nucl-ex/0309004].

\bibitem{Fukuda:1998mi}
  Y.~Fukuda {\it et al.}  [Super-Kamiokande Collaboration],
  Phys.\ Rev.\ Lett.\  {\bf 81} (1998) 1562
  [arXiv:hep-ex/9807003].

\bibitem{Ambrosio:2001je}
  M.~Ambrosio {\it et al.}  [MACRO Collaboration],
  Phys.\ Lett.\  B {\bf 517} (2001) 59
  [arXiv:hep-ex/0106049].
  
\bibitem{Apollonio:1999ae}
  M.~Apollonio {\it et al.}  [CHOOZ Collaboration],
  Phys.\ Lett.\  B {\bf 466} (1999) 415
  [arXiv:hep-ex/9907037].

\bibitem{Apollonio:2002gd}
  M.~Apollonio {\it et al.}  [CHOOZ Collaboration],
  Eur.\ Phys.\ J.\  C {\bf 27} (2003) 331
  [arXiv:hep-ex/0301017].
  
\bibitem{Boehm:2001ik}
  F.~Boehm {\it et al.},
  Phys.\ Rev.\  D {\bf 64} (2001) 112001
  [arXiv:hep-ex/0107009].

\bibitem{Eguchi:2002dm}
  K.~Eguchi {\it et al.}  [KamLAND Collaboration],
  Phys.\ Rev.\ Lett.\  {\bf 90} (2003) 021802
  [arXiv:hep-ex/0212021].

\bibitem{Ahn:2002up}
  M.~H.~Ahn {\it et al.}  [K2K Collaboration],
  Phys.\ Rev.\ Lett.\  {\bf 90} (2003) 041801
  [arXiv:hep-ex/0212007].

\bibitem{Aliu:2004sq}
  E.~Aliu {\it et al.}  [K2K Collaboration],
  Phys.\ Rev.\ Lett.\  {\bf 94} (2005) 081802
  [arXiv:hep-ex/0411038].

\bibitem{Michael:2006rx}
  D.~G.~Michael {\it et al.}  [MINOS Collaboration],
  Phys.\ Rev.\ Lett.\  {\bf 97} (2006) 191801
  [arXiv:hep-ex/0607088].

\bibitem{Pontecorvo:1957cp}
  B.~Pontecorvo,
  Sov.\ Phys.\ JETP {\bf 6} (1957) 429
  [Zh.\ Eksp.\ Teor.\ Fiz.\  {\bf 33} (1957) 549].

\bibitem{Maki:1962mu}
  Z.~Maki, M.~Nakagawa and S.~Sakata,
  Prog.\ Theor.\ Phys.\  {\bf 28} (1962) 870.

\bibitem{Pontecorvo:1967fh}
  B.~Pontecorvo,
  Sov.\ Phys.\ JETP {\bf 26} (1968) 984
  [Zh.\ Eksp.\ Teor.\ Fiz.\  {\bf 53} (1967) 1717].

\bibitem{Gribov:1968kq}
  V.~N.~Gribov and B.~Pontecorvo,
  Phys.\ Lett.\  B {\bf 28} (1969) 493.

\bibitem{GonzalezGarcia:2007ib}
  M.~C.~Gonzalez-Garcia and M.~Maltoni,
  arXiv:0704.1800 [hep-ph].

\bibitem{boh}
  C.~Aalseth {\it et al.},
  [arXiv:hep-ph/0412300];
  J.~Lesgourgues and S.~Pastor,
  Phys.\ Rept.\  {\bf 429} (2006) 307
  [arXiv:astro-ph/0603494];
  S.~Hannestad,
  arXiv:0710.1952 [hep-ph].

\bibitem{Cervera:2000kp}
  A.~Cervera, A.~Donini, M.~B.~Gavela, J.~J.~Gomez Cadenas, P.~Hernandez, O.~Mena and S.~Rigolin,
  Nucl.\ Phys.\ B {\bf 579} (2000) 17 [Erratum-ibid.\ B {\bf 593} (2001) 731] [arXiv:hep-ph/0002108].

\bibitem{Donini:2002rm}
  A.~Donini, D.~Meloni and P.~Migliozzi,
  Nucl.\ Phys.\ B {\bf 646} (2002) 321 [arXiv:hep-ph/0206034];
  J.\ Phys.\ G {\bf 29} (2003) 1865 [arXiv:hep-ph/0209240];
  D.~Autiero {\it et al.},
  Eur.\ Phys.\ J.\ C {\bf 33} (2004) 243
  [arXiv:hep-ph/0305185].

\bibitem{BurguetCastell:2001ez}
  J.~Burguet-Castell, M.~B.~Gavela, J.~J.~Gomez-Cadenas, P.~Hernandez and O.~Mena,
  Nucl.\ Phys.\ B {\bf 608} (2001) 301 [arXiv:hep-ph/0103258].

\bibitem{Minakata:2001qm}
  H.~Minakata and H.~Nunokawa,
  JHEP {\bf 0110} (2001) 001 [arXiv:hep-ph/0108085].

\bibitem{Fogli:1996pv}
  G.~L.~Fogli and E.~Lisi,
  Phys.\ Rev.\ D {\bf 54} (1996) 3667 [arXiv:hep-ph/9604415].

\bibitem{Barger:2001yr}
  V.~Barger, D.~Marfatia and K.~Whisnant,
  Phys.\ Rev.\ D {\bf 65} (2002) 073023 [arXiv:hep-ph/0112119].

\bibitem{Donini:2005rn}
  A.~Donini, D.~Meloni and S.~Rigolin,
  Eur.\ Phys.\ J.\ C {\bf 45} (2006) 73
  [arXiv:hep-ph/0506100].
  D.~Meloni,
  Nucl.\ Phys.\ Proc.\ Suppl.\  {\bf 155}, 178 (2006)
  [arXiv:hep-ph/0509370].

\bibitem{Ardellier:2006mn}
  F.~Ardellier {\it et al.}  [Double Chooz Collaboration],
  [arXiv:hep-ex/0606025].

\bibitem{Itow:2001ee}
  Y.~Itow {\it et al.},
  [arXiv:hep-ex/0106019].

\bibitem{Zucchelli:sa}
  P.~Zucchelli,
  Phys.\ Lett.\  B {\bf 532} (2002) 166.

\bibitem{Geer:1997iz}
  S.~Geer,
  Phys.\ Rev.\ D {\bf 57} (1998) 6989 [Erratum-ibid.\ D {\bf 59} (1999) 039903] [arXiv:hep-ph/9712290];
  A.~De Rujula, M.~B.~Gavela and P.~Hernandez,
  Nucl.\ Phys.\ B {\bf 547} (1999) 21 [arXiv:hep-ph/9811390].

\bibitem{Apollonio:2002en}
  M.~Apollonio {\it et al.},
  [arXiv:hep-ph/0210192].

\bibitem{Group:2007kx}
  The ISS Physics Working Group,
  arXiv:0710.4947 [hep-ph].

\bibitem{Cervera:2000vy}
A.~Cervera, F.~Dydak and J.~Gomez Cadenas,
Nucl.\ Instrum.\ Meth.\ A {\bf 451} (2000) 123.

\bibitem{Burguet-Castell:2003vv}
  J.~Burguet-Castell, D.~Casper, J.~J.~Gomez-Cadenas, P.~Hernandez and F.~Sanchez,
  Nucl.\ Phys.\ B {\bf 695}, 217 (2004) [arXiv:hep-ph/0312068].

\bibitem{Burguet-Castell:2005va}
  J.~Burguet-Castell, D.~Casper, E.~Couce, J.~J.~Gomez-Cadenas and P.~Hernandez,
  Nucl.\ Phys.\ B {\bf 725}, 306 (2005)
  [arXiv:hep-ph/0503021].

\bibitem{GilBotella:2007uv}
  I.~Gil-Botella  [Double Chooz Collaboration],
  arXiv:0710.4258 [hep-ex].

\bibitem{Mena:2006uw}
  O.~Mena, H.~Nunokawa and S.~J.~Parke,
  Phys.\ Rev.\  D {\bf 75} (2007) 033002
  [arXiv:hep-ph/0609011].

\bibitem{Huber:2002mx}
  P.~Huber, M.~Lindner and W.~Winter,
  Nucl.\ Phys.\  B {\bf 645} (2002) 3
  [arXiv:hep-ph/0204352].

\bibitem{Huber:2006wb}
  P.~Huber, M.~Lindner, M.~Rolinec and W.~Winter,
  Phys.\ Rev.\  D {\bf 74} (2006) 073003
  [arXiv:hep-ph/0606119].

\bibitem{Huber:2003ak}
  P.~Huber and W.~Winter,
  Phys.\ Rev.\  D {\bf 68} (2003) 037301
  [arXiv:hep-ph/0301257].

\bibitem{Bueno:2000fg}
  A.~Bueno, M.~Campanelli and A.~Rubbia,
  Nucl.\ Phys.\  B {\bf 589} (2000) 577
  [arXiv:hep-ph/0005007].

\bibitem{cerveragolden07}
  T.~Abe {\it et al.},
  arXiv:0712.4129 [physics.ins-det].

\bibitem{MERIT}
MERIT Homepage
http://proj-hiptarget.web.cern.ch/proj-hiptarget/

\bibitem{MICE}
  D.~M.~Kaplan and K.~Long,
  arXiv:0707.1915 [physics.acc-ph].

\bibitem{Bouchez:2003fy}
  J.~Bouchez, M.~Lindroos and M.~Mezzetto,
  AIP Conf.\ Proc.\  {\bf 721} (2004) 37 [arXiv:hep-ex/0310059];
  M.~Mezzetto,
  J.\ Phys.\ G {\bf 29} (2003) 1771 [arXiv:hep-ex/0302007];
  Nucl.\ Phys.\ Proc.\ Suppl.\  {\bf 143} (2005) 309 [arXiv:hep-ex/0410083].

\bibitem{Donini:2004hu}
  A.~Donini, E.~Fern\'andez-Mart\'{\i}nez, P.~Migliozzi, S.~Rigolin and L.~Scotto Lavina,
  Nucl.\ Phys.\ B {\bf 710} (2005) 402 [arXiv:hep-ph/0406132].

\bibitem{Donini:2004iv}
  A.~Donini, E.~Fern\'andez-Mart\'{\i}nez and S.~Rigolin,
  Phys.\ Lett.\ B {\bf 621} (2005) 276 [arXiv:hep-ph/0411402].

\bibitem{PAF}
  ``Proton Accelerator for the Future'' (PAF) inter-departmental working group webpage, http://pofpa.web.cern.ch/pofpa/

\bibitem{Jansson:2007nm}
  A.~Jansson, O.~Mena, S.~Parke and N.~Saoulidou,
  arXiv:0711.1075 [hep-ph].

\bibitem{Rubbia:2006pi}
  C.~Rubbia, A.~Ferrari, Y.~Kadi and V.~Vlachoudis,
  Nucl.\ Instrum.\ Meth.\  A {\bf 568} (2006) 475
  [arXiv:hep-ph/0602032].

\bibitem{Rubbia:2006zv}
  C.~Rubbia,
  [arXiv:hep-ph/0609235].

\bibitem{Euronu}
``EURO-$\nu$ High intensity $\nu$ ocillation facility in Europe'',
FP7-infrastructures-2007-1 project number 212372.  

\bibitem{Donini:2006dx}
  A.~Donini and E.~Fern\'andez-Mart\'{\i}nez,
  Phys.\ Lett.\  B {\bf 641} (2006) 432
  [arXiv:hep-ph/0603261].

\bibitem{Donini:2006tt}
  A.~Donini, E.~Fern\'andez-Mart\'{\i}nez, P.~Migliozzi, S.~Rigolin, L.~Scotto Lavina, T.~Tabarelli de Fatis and F.~Terranova,
  Eur.\ Phys.\ J.\  C {\bf 48}, 787 (2006)
  [arXiv:hep-ph/0604229].

\bibitem{Donini:2007qt}
  A.~Donini {\it et al.},
  Eur.\ Phys.\ J.\  C {\bf 53} (2008) 599
  [arXiv:hep-ph/0703209].

\bibitem{Agarwalla:2006vf}
  S.~K.~Agarwalla, S.~Choubey and A.~Raychaudhuri,
  Nucl.\ Phys.\  B {\bf 771} (2007) 1
  [arXiv:hep-ph/0610333].

\bibitem{Agarwalla:2007ai}
  S.~K.~Agarwalla, S.~Choubey and A.~Raychaudhuri,
  arXiv:0711.1459 [hep-ph].

\bibitem{ino}
  G.~Rajasekaran,
  AIP Conf.\ Proc.\  {\bf 721} (2004) 243
  [arXiv:hep-ph/0402246].

\bibitem{Agarwalla:2006gz}
  S.~K.~Agarwalla, S.~Choubey, S.~Goswami and A.~Raychaudhuri,
  Phys.\ Rev.\  D {\bf 75} (2007) 097302
  [arXiv:hep-ph/0611233].

\bibitem{Autin:2002ms}
  B.~Autin {\it et al.},
  J.\ Phys.\ G {\bf 29} (2003) 1785
  [arXiv:physics/0306106].

\bibitem{betadecays}
  L.P. Ekström and R.B. Firestone, WWW Table of Radioactive Isotopes, \\
  database version 2/28/99 from URL http://ie.lbl.gov/toi/index.htm

\bibitem{Tengblad}
  O. Tengblad, private communication, \\
  http://www.targisol.csic.es/intro\_database.html

\bibitem{Volpe:2003fi}
  C.~Volpe,
  J.\ Phys.\ G {\bf 30}, L1 (2004)
  [arXiv:hep-ph/0303222];
  J.\ Phys.\ G {\bf 34} (2007) R1
  [arXiv:hep-ph/0605033];
  R.~Lazauskas, A.~B.~Balantekin, J.~H.~De Jesus and C.~Volpe,
  Phys.\ Rev.\  D {\bf 76} (2007) 053006
  [arXiv:hep-ph/0703063].

\bibitem{Campagne:2006yx}
  J.~E.~Campagne, M.~Maltoni, M.~Mezzetto and T.~Schwetz,
  JHEP {\bf 0704} (2007) 003 [arXiv:hep-ph/0603172].

\bibitem{Huber:2005jk}
  P.~Huber, M.~Lindner, M.~Rolinec and W.~Winter,
  Phys.\ Rev.\ D {\bf 73}, 053002 (2006) [arXiv:hep-ph/0506237].

\bibitem{Terranova:2004hu}
  F.~Terranova, A.~Marotta, P.~Migliozzi and M.~Spinetti,
  Eur.\ Phys.\ J.\ C {\bf 38}, 69 (2004) [arXiv:hep-ph/0405081].

\bibitem{bruning}
  O.~Bruning {\it et al.}, ``LHC luminosity and energy upgrade: A
  feasibility study,'' CERN-LHC-PROJECT-REPORT-626, 2002.

\bibitem{Lindroos2}
See the talk by M. Lindroos at the second plenary IDS meeting. \\
http://www.hep.ph.ic.ac.uk/ids/communication/RAL-2008-01-16/agenda.html

\bibitem{lipari}
  P.~Lipari, private communication;
  P.~Lipari, M.~Lusignoli and F.~Sartogo,
  Phys.\ Rev.\ Lett.\  {\bf 74}, 4384 (1995) [arXiv:hep-ph/9411341].

\bibitem{nostrareview}
A.~Guglielmi, M.~Mezzetto, P.~Migliozzi and F.~Terranova, [arXiv:hep-ph/0508034].

\bibitem{monocromatic}
  J.~Sato,
  Phys.\ Rev.\ Lett.\  {\bf 95}, 131804 (2005)
  [arXiv:hep-ph/0503144];
  J.~Bernabeu, J.~Burguet-Castell, C.~Espinoza and M.~Lindroos,
  JHEP {\bf 0512}, 014 (2005)
  [arXiv:hep-ph/0505054];
  M.~Rolinec and J.~Sato,
  JHEP {\bf 0708} (2007) 079
  [arXiv:hep-ph/0612148].

\bibitem{monolith}
  N.~Y.~Agafonova {\it et al.}  [MONOLITH Collaboration],
  LNGS-P26-2000, CERN-SPSC-2000-031, CERN-SPSC-M-657.

\bibitem{Donini:2005db}
  A.~Donini, E.~Fern\'andez-Mart\'{\i}nez, D.~Meloni and S.~Rigolin,
  Nucl.\ Phys.\  B {\bf 743}, 41 (2006)
  [arXiv:hep-ph/0512038].

\bibitem{Mezzetto:2003ub}
  M.~Mezzetto,
  J.\ Phys.\ G {\bf 29}, 1771 (2003)
  [arXiv:hep-ex/0302007].

\bibitem{Ashie:2005ik}
  Y.~Ashie {\it et al.}  [Super-Kamiokande Collaboration],
  Phys.\ Rev.\  D {\bf 71} (2005) 112005
  [arXiv:hep-ex/0501064].

\bibitem{Globes}
  P.~Huber, J.~Kopp, M.~Lindner, M.~Rolinec and W.~Winter,
  Comput.\ Phys.\ Commun.\  {\bf 177} (2007) 432
  [arXiv:hep-ph/0701187].

\bibitem{Choubey:2005zy}
  S.~Choubey and P.~Roy,
  Phys.\ Rev.\  D {\bf 73} (2006) 013006
  [arXiv:hep-ph/0509197].

\bibitem{Huber:2005ep}
  P.~Huber, M.~Maltoni and T.~Schwetz,
  Phys.\ Rev.\  D {\bf 71} (2005) 053006
  [arXiv:hep-ph/0501037].

\bibitem{betabeams_moriond} 
  Workshop on ``Radioactive beams for nuclear physics and neutrino physics''
  37$^{th}$ Rencontre de Moriond, Les Arcs (France) March 17-22nd, 2003;
  http://moriond.in2p3.fr/radio/index.html.

\bibitem{garoby_hif04}
  R.~Garoby and W.~Scandale,
  Nucl.\ Phys.\ Proc.\ Suppl.\  {\bf 147} (2005) 16.

\bibitem{Lindroos}
  M. Lindroos, http://beta-beam-parameters.web.cern.ch/beta-beam-parameters/

\bibitem{Wildner:2007dy}
  E.~Wildner, M.~Benedikt, N.~Emelianenko, A.~Fabich, S.~Hancock and M.~Lindroos,
  CERN-AB-2007-015, CERN-AT-2007-016.

\bibitem{SPL}
  B.~Autin {\it et al.},
  CERN-2000-012

\bibitem{Garoby:2005se} 
  R.~Garoby, ``The SPL at CERN'', CERN-AB-2005-007, contribution to the ``33rd ICFA Advanced Beam
  Dynamics Workshop: High Intensity High Brightness Hadron Beams (ICFA HB2004)'', Bensheim, Germany, 2004.

\bibitem{Campagne:2004wt}
  J.~E.~Campagne and A.~Cazes,
  Eur.\ Phys.\ J.\  C {\bf 45} (2006) 643
  [arXiv:hep-ex/0411062].

\bibitem{Mezzetto:2003mm}
  M.~Mezzetto,
  J.\ Phys.\ G {\bf 29}, 1781 (2003)
  [arXiv:hep-ex/0302005].

\bibitem{fabbricatore}
  P.~Fabbricatore, S.~Farinon, M.~Greco, U.~Gambardella and G.~Volpini,
  Nucl.\ Phys.\ Proc.\ Suppl.\  {\bf 154} (2006) 157.

\bibitem{terranova_nufact04}
  F.~Terranova,
  Nucl.\ Phys.\ Proc.\ Suppl.\  {\bf 149}, 185 (2005).

\bibitem{Devred:2005vd}
  A.~Devred {\it et al.},
  ``Status of the Next European Dipole (NED) activity of the Collaborated
  Accelerator Research in Europe (CARE) project,''
  CERN-AT-2005-002.

\bibitem{FabichNuFact07}
A.~Fabich, talk at Nufact07.

\bibitem{Winter:2003ac}
  W.~T.~Winter {\it et al.},
  Phys.\ Rev.\ Lett.\  {\bf 91} (2003) 252501
  [Nucl.\ Phys.\ A {\bf 746} (2004) 311].

\end{thebibliography}
\end{document}